# Nematic twist-bend phase of a bent liquid crystal dimer: field-induced deformations of the helical structure, macroscopic polarization and fast switching speeds†

Sourav Patranabish[a], Aloka Sinha[a]*, Madhu B. Kanakala[b] and C. V. Yelamaggad[b]

[a] *Department of Physics, Indian Institute of Technology Delhi, Hauz Khas, New Delhi 110016, India;*

[b] *Centre for Nano and Soft Matter Sciences, P. B. No. 1329, Prof. U. R. Rao Road, Jalahalli, Bengaluru 560013, India;*

**Corresponding author: aloka@physics.iitd.ac.in; alokaphysics@gmail.com*

† *Electronic Supplementary Information (ESI) available*

# Nematic twist-bend phase of a bent liquid crystal dimer: field-induced deformations of the helical structure, macroscopic polarization and fast switching speeds

The twist-bend nematic ($N_{tb}$) phase is a recent addition to the nematic (N) phases of liquid crystals (LCs). A net polar order in the $N_{tb}$ phase under an external electric field was predicted in several recent theoretical studies but yet to be experimentally realized. We investigated the polar nature, dielectric properties, electro-optical switching and optical transmission properties of a bent LC dimer CB7CB. The LC showed a relatively high-temperature nematic (N) phase and a lower-temperature nematic ($N_X$) phase (also called $N_{tb}$ in literature). A threshold-dependent polarization current response with large polarization values was obtained in the entire mesophase range. The associated switching times were found in sub-millisecond region. This ferroelectric-like polarization resulted from collective reorientation of polar cybotactic clusters. In $N_X$ phase, electric field-induced deformation of twisted helical structures also contributed to net polarization. Dielectric measurements confirmed the presence of cybotactic clusters *via* relaxation processes with large activation energies. Deformation of the $N_X$ helical structure under external electric field was corroborated by polarized optical microscopy and optical transmission studies. The field-induced deformations, net polar order and fast switching will contribute towards greater understanding of the $N_X$ (or $N_{tb}$) phase dynamics. It may also find applications in next-generation electro-optic devices.

## 1. Introduction

The liquid crystals (LCs) comprised of bent-shaped molecules (Bent-core LCs or BLCs) are often considered as the material for next-generation displays and electro-optic devices [1-3]. They differ from their calamitic counterpart primarily in terms of the molecular shape and symmetry [4]. The nematic (N) phase of BLCs possesses several unique and interesting properties such as polarity, biaxiality and chirality, with potential usage as the material for fast switching displays [5-8]. Owing to these remarkable possibilities, they have been a subject of intense experimental and theoretical research in recent years [1,2]. The typical switching time for the nematic LCs used in present-day displays and electro-optic devices is in the range of milliseconds. An underlying polar nature in the N phase of BLCs is expected to provide with much faster switching [9,10], preferably in the sub-millisecond region. Although predicted, such a fast switching in the N phase is yet to be experimentally realized.

The latest inclusion in the nematic phase sequence of BLCs is the twist-bend nematic ($N_{tb}$) phase. It was first proposed by R. B Meyer in 1973 in the nematic phases with spontaneous polarisation (*i.e.* in a ferroelectric nematic) [11,12]. Meyer proposed that for bent-shaped molecules, a coupling between the ferroelectric and the flexoelectric polarisations will lead to a heliconical structure, which is now referred to as the $N_{tb}$ phase in literature [11,12]. Following this, in 1999, Lorman and Mettout predicted a new mesophase for curved molecules with a transverse circular polar order which is similar to the Meyer's $N_{tb}$ phase [12,13]. In 2001, in an alternative approach, Dozov showed that the twist deformation in the $N_{tb}$ phase of bent-shaped molecules is not spontaneous, but it is imposed by the spontaneous bend deformation, enabling a 'third dimension escape' for $\hat{\boldsymbol{n}}$ [14]. In the past few decades, several such models for the $N_{tb}$ phase have been proposed based on both polar [13,15-17] and apolar [14,18-25] interactions [12]. In fact, there is not a unique $N_{tb}$ model, but several different models proposed and developed independently by many researchers [12]. The common feature in all of these models is that in the $N_{tb}$ phase, the nematic director ($\hat{\boldsymbol{n}}$) draws an oblique helicoid maintaining a constant oblique angle $\theta$ with the z-axis (the helix axis), such that $\mathbf{0} < \boldsymbol{\theta} < \boldsymbol{\pi}/\mathbf{2}$. The $N_{tb}$ phase is thus chiral, even though it is constituted of achiral mesogens [26]. While theoretical predictions of the $N_{tb}$ phase were made long ago, experimental realizations were made only recently [26,27]. Cestari *et al.*, in their pioneering paper [27], first reported the experimental identification of this phase in a bent LC dimer CB7CB which is the most studied dimer to exhibit this phase. At that time they marked this phase as 'X' (or $N_X$, as identified in the literature [12]) and presented with a detailed experimental investigation, which was subsequently established as the $N_{tb}$ phase [26,27]. The value of the bend elastic constant ($K_{33}$) in $N_{tb}$ phase was predicted to be negative because of the spontaneous bend deformation [14]. Experimental studies, on the contrary, have demonstrated only a small (non-negative) value of $K_{33}$ [26,28].

The $N_{tb}$ phase is an example of spontaneous chiral symmetry breaking in liquid crystals, reminiscent of the helical smectic C* (SmC*) phase [15,24]. Due to chirality and the spontaneous breaking of mirror symmetry, this phase is expected to be locally polar and may also exhibit the heli-electric nature [12-14,24,29]. Dozov, through his studies, indicated the possibility of large spontaneous electric polarization in the $N_{tb}$ phase [14]. This idea was further supported by the work of Shamid *et al.*, where they suggest that this modulated polar phase is locally ferroelectric and have a spontaneous polar order that leads to electrostatic polarization [15]. In the $N_{tb}$ phase, the local polar order (pointing along the bend-direction) is modulated in a helix and hence averages to zero globally, similar to the ferroelectric SmC* LCs [15,29]. According to the ferroelectric models, the spontaneous polarisation exists even in the absence of the director distortion and a strong electric field, perpendicular to the helix axis, is expected to completely (or perhaps only partially) unwind, the helical structure of the $N_{tb}$ phase [12,25,30]. Therefore, in the $N_{tb}$ phase, the helical structure needs to be deformed or unwound to obtain macroscopic polar behaviour. Although predicted in several theoretical studies, the experimental identification of ferroelectric polarization in the $N_{tb}$ phase remains elusive. The unwinding or deformation of helical structure to obtain a macroscopic polarization is an established technique in ferroelectric liquid crystals (FLCs). It has been widely utilized in surface stabilized FLCs (SSFLCs), deformed helix FLCs and vFLCs, to name a few [31-33]. In the $N_{tb}$ phase, only a single experimental report exists showing signatures of a polar order, where the authors report a linear polar switching under an applied electric field in homologous CBnCB (n = 9, 11) compounds [34]. A flexoelectrically driven electroclinic effect with response times in sub-microsecond time-scales has also been reported in the bent flexible dimer CB7CB [25]. Pajak *et al.*, in a recent theoretical investigation, have demonstrated that under a sufficiently large electric field, the twist-bend helix can be unwound giving rise to a globally polar nematic phase ($N_P$) with non-vanishing polarization [24]. Similar predictions were also made by Merkel *et al.*, where they report flexoelectric polarization and electroclinic effect arising due to this polarization [35]. Through dielectric studies, they also report low-frequency collective relaxations and a large polar order (dielectric strength ~ 800) in the $N_{tb}$ phase. A net polar order in the $N_{tb}$ phase is fascinating since it is new and may find application in fast-switching electro-optic devices. A detailed and rigorous investigation of the N and $N_{tb}$ phases of BLCs is therefore essential for gaining significant insight into the polar nature and interesting switching behaviour.

Recently, another group of researchers have independently developed a model to describe this new and exciting phase discovered experimentally by Cestari *et al.* [27,36-42]. They argue that the $N_X$ phase discovered by Cestari *et al.* has been incorrectly identified as the $N_{tb}$ phase proposed by Meyer and justify that the $N_X$ phase is actually a polar-twisted nematic ($N_{PT}$) phase [40]. The main argument behind their claims is the short pitch exhibited by the $N_X$ phase (~ 10 nm) which is several orders of magnitude smaller than the values predicted by Meyer (~ 1 μm or higher) [12,36-41]. They present detailed experimental and theoretical investigations of the bent LC dimer CB7CB and its higher homologues in support of their claims and identify the $N_X$ phase as the polar-twisted nematic ($N_{PT}$) phase. In one of their recent papers, they also discuss the nature of the high-temperature uniaxial nematic ($N_U$) phase exhibited by CB7CB [41]. The authors discuss that this high-temperature nematic (N) is a cybotactic nematic phase consisting of $N_{PT}$ clusters ($N_{CybPT}$), supported by several different observations [41]. They further discuss that the clusters present in the high-temperature N phase have the same structure as the polar-twisted nematic ($N_{PT}$) domains but with a much smaller domain-size. These results are consistent with the findings reported by Krishnamurthy *et al.*, where they report clusters with a helical internal structure being present in the high-temperature N phase [43]. The cybotactic clusters exhibited by the majority of BLCs are locally polar (with large transverse dipole moments), and they are known to give rise to a net ferroelectric polarization under an external electric field above a certain threshold [1,3,44-56]. Also, the CB7CB dimer has a large transverse dipole moment, and the helical structures exhibited by the dimer at lower temperatures are locally polar [12,14,24,29]. Therefore, the clusters present in the high-temperature N phase are expected to be locally polar. Under a strong external electric field – (i) in the high-temperature N phase, a net polar response is expected due to reorientation of the polar clusters, and (ii) in the $N_X$ phase, a net polar order is expected due to unwinding of the helical structure (complete or partial unwinding), under the condition E (electric field) ⊥ h (helix axis) [12,25,30]. Experimentally, these findings, if obtained, will provide strong grounds to the long-predicted twist-bend phase by Meyer with a transverse ferroelectric polarization. Therefore, experimental and theoretical studies are required in this direction for unequivocal identification of the nematic phases exhibited by the LC dimer CB7CB. To avoid any contradiction with the identification or nomenclature of the low-temperature nematic phase of CB7CB, we will denote it as the $N_X$ phase in the rest of this manuscript, unless otherwise specified.

In this paper, we have investigated the polarization and switching behaviour of a bent LC dimer CB7CB (1",7"-bis(4-cyanobiphenyl-4'-yl)heptane) in the nematic (N) and the low-temperature nematic ($N_X$) phases. The dielectric, optical and electro-optic measurements were performed in the various mesophases. We report a net polarization in the N and $N_X$ phases of the LC. Through dielectric and electro-optic experimental results, we discuss the origin of this polar order. Optical birefringence and order parameter measurements were performed and the effect of externally applied electric field on these parameters have been demonstrated. Frequency-dependent flexoelectric effects arising in the field-on state are also discussed. The study aims to establish a correlation between the experimental results and the recent theoretical predictions by several researchers of a polar order in the N and $N_X$ phases of the LC CB7CB. Other details are available under Supplementary Information.

## 2. Experimental details

### *2.1. Materials*

The LC material 1”,7”-bis(4-cyanobiphenyl-4’-yl)heptane (CB7CB) used in this study was synthesized at the Centre for Nano and Soft Matter Sciences (CeNS), Bengaluru, India. The chemical formula of the LC compound is shown in Figure 1 and the related synthetic scheme details are available under the electronic supplementary information (ESI). The phase transition temperatures of the compound were determined using a Perkin-Elmer Diamond differential scanning calorimetry (DSC), and also using an Olympus BX-51P polarizing optical microscope (POM). The DSC measurements were taken at a heating and cooling rate of 5 °C/min. Before use, the DSC was calibrated using pure indium as a standard. The POM measurements were taken only in the cooling cycle and at a lower rate of cooling (0.5 °C/min). The obtained transition temperatures are summarized in Table 1. The apparent difference between the obtained transition temperatures is possibly because of the difference in the rate of cooling.

Figure 1. Chemical formula of the twist-bend nematic LC compound CB7CB (1",7"-bis(4-cyanobiphenyl-4'-yl)heptane).

**Table 1**. Phase sequence observed in this study (using POM & DSC) and that reported by other researchers [26,29,62-65]

| Compound CB7CB | Heating | Cooling |
|---|---|---|
| **V. Borshch *et al.* [26]** | X 84.5 $N_{tb}$ 103.3 N 116.6 I | I 116 N 103 $N_{tb}$ 83 X |
| **G. Babakhanova *et al.* [62]** | ----- | I 114 N 101 $N_{tb}$ |
| **C. Meyer *et al.* [63]** | ----- | I 114.6 N 100.8 $N_{tb}$ |
| **N. Vaupotic *et al.* [64]** | ----- | I 114.3 N 101.6 $N_{tb}$ |
| **D. Chen *et al.* [29]** | $N_{tb}$ 100 N 113 I | I 112 N 99 $N_{tb}$ |
| **C. Meyer *et al.* [65]** | ----- | I 115.8 N 103.1 $N_{tb}$ |
| **DSC** | Cr 104.09 N 113.83 I | I 112.64 N 100.51 $N_X$ |
| **POM observation** | ---- | I 115.5 N 103 $N_X$ |

### *2.2. Methods*

#### *2.2.1. Optical textures*

Indium Tin Oxide (ITO) coated planar cells (Instec Inc., USA) of thickness 5 μm were used for the experiments. The LC material was filled in the LC cells *via.* capillary action around 10 °C above the isotropic-nematic transition temperature. The temperature was maintained with a temperature controller (Instec. MK1000) attached to a hot-stage (Instec. HCS302). The optical textures of the LC compound were recorded between crossed polarizers using an OLYMPUS BX-51P polarizing optical microscope (POM). All the measurements were carried out while slowly cooling the sample from the isotropic phase.

#### *2.2.2. Electro-optic measurements*

The electro-optical measurements were carried out by using planar LC cells of thickness 5 μm and 3.2 μm with active electrode area 1 $cm^2$ and 0.25 $cm^2$, respectively. The polar switching was studied by analysing the current response across a 25 kΩ resistance using a Tektronix AFG3021 function generator, a TPS 2024 digital oscilloscope and a homebuilt amplifier of gain 10, FWHM 0-10kHz. The LC was probed with an external electric field of amplitude up to 18 $V_{PP}$/μm and frequencies ranging from 1 Hz to around 150 Hz. Both triangular and square-wave input signals were used for the measurements. The values of polarization (*P*) were calculated from the current response using the repolarization current measurement technique (triangular-wave method) [44]. The relation used to determine *P* is, $\boldsymbol{P} = \frac{\boldsymbol{1}}{\boldsymbol{2R_S A}} \int \boldsymbol{\Delta V dt}$, where $R_S$ is the resistance (25 kΩ) connected in series with the cell, *A* is the cell's active area and $\Delta V$ is the excess voltage due to the polarization reversal.

*2.2.3. Dielectric measurements*

For the dielectric spectroscopy measurements, an Agilent E4980A precision LCR meter was used in the frequency range 20 Hz – 2 MHz with a measuring voltage of 0.1 V ($V_{rms}$). The frequency-dependent complex dielectric permittivity is given by, $\boldsymbol{\varepsilon^*(f) = \varepsilon'(f) - i\varepsilon''(f)}$, where $\varepsilon'(f)$ and $\varepsilon''(f)$ represent the real and the imaginary parts of the complex permittivity, respectively. The characteristic dielectric parameters such as the relaxation frequency ($f_R$) and the dielectric strength ($\delta\varepsilon$) were obtained by fitting the dielectric loss ($\varepsilon''$) data to the imaginary part of the well-known Havriliak-Negami (H-N) fit function [44,57-61],

$$\boldsymbol{\varepsilon'' = \frac{\sigma_0}{\varepsilon_0(2\pi f)^s} + \sum_{k=1}^{N} \mathrm{Im}\left\{\frac{\delta\varepsilon_k}{[1+(i2\pi f\tau_k)^{\alpha_k}]^{\beta_k}}\right\}}$$

$$\boldsymbol{= \frac{\sigma_0}{\varepsilon_0(2\pi f)^s} + \sum_{k=1}^{N} \frac{\delta\varepsilon_k \sin(\beta_k\theta)}{[1+(2\pi f\tau_k)^{2\alpha_k} + 2(2\pi f\tau_k)^{\alpha_k}\cos(\alpha_k\pi/2)]^{\beta_k/2}}} \qquad (1)$$

where,

$$\boldsymbol{\theta = tan^{-1}\left[\frac{(2\pi f\tau_k)^{\alpha_k}\, sin(\alpha_k\pi/2)}{1+(2\pi f\tau_k)^{\alpha_k}\, cos(\alpha_k\pi/2)}\right]}$$

Here, $f$ is the frequency, $\boldsymbol{\varepsilon_\infty}$ is the high-frequency limit of permittivity ($\varepsilon'$), $\boldsymbol{\delta\varepsilon_k}$ is the dielectric strength, $\boldsymbol{\sigma_0}$ is the dc conductivity, $\boldsymbol{\varepsilon_0}$ is the free-space permittivity ($8.854*10^{-12}$ F/m). $s$ is a fitting parameter responsible for the nonlinearity in dc conductivity part (for ohmic behaviour, $s = 1$), $k$ is the number of relaxation processes, $\boldsymbol{\tau_k}$ (= $\boldsymbol{1/2\pi f_k}$) is the relaxation time for $k$-th relaxation process. $\boldsymbol{\alpha_k}$ and $\boldsymbol{\beta_k}$ are empirical fit parameters that describe symmetric and non-symmetric broadening, respectively, of the $k$-th relaxation peak. For static dielectric measurements, signals of 20 V, 1kHz and 0.1 V, 1kHz were applied to measure $\boldsymbol{\varepsilon}_{||}$ and $\boldsymbol{\varepsilon}_{\perp}$, respectively. The dielectric anisotropy (Δε) was computed using the relation, $\Delta\boldsymbol{\varepsilon} = \boldsymbol{\varepsilon}_{||} - \boldsymbol{\varepsilon}_{\perp}$.

*2.2.4. Optical transmission studies*

For optical transmission measurements, the planar LC sample was placed between two crossed (P⊥A) Glan-Thompson polarizers (GTH10M, Thorlabs, Inc.) and it was perpendicularly illuminated with a He-Ne Laser (~ 633 nm). The rubbing direction ($\vec{\boldsymbol{r}}$) (*i.e.* the LC director ($\hat{\boldsymbol{n}}$)) of the planar cell was set at 45° with respect to the crossed polarizer (or analyzer) pass-axis, for birefringence (Δ*n*) measurements. The transmitted power at the output end was measured using a Gentec PH100-Si-HA-OD1 photo-detector attached to a Gentec Maestro power meter. The birefringence (Δ*n*) was then evaluated using the relation [66],

$$\boldsymbol{I} = \boldsymbol{I_0}\, \boldsymbol{sin^2 2\varphi}\, \boldsymbol{sin^2}\frac{\boldsymbol{\delta}}{\boldsymbol{2}} = \frac{\boldsymbol{I_0}}{\boldsymbol{2}}\,(\boldsymbol{1} - \boldsymbol{cos\delta}) \tag{2}$$

Here, $\boldsymbol{\delta} = \frac{\boldsymbol{2\pi}}{\boldsymbol{\lambda}}\Delta\boldsymbol{nd}$ is the phase difference, *I* is the transmitted light intensity, $I_0$ is the incident light intensity, *φ* (= 45°) is the azimuthal angle/angle made by the LC director ($\hat{\boldsymbol{n}}$) with the polarizer/analyzer pass axis, λ is the incident light wavelength, $\Delta\boldsymbol{n} = \boldsymbol{n_e} - \boldsymbol{n_o}$ is the birefringence, $n_e$ and $n_o$ are the extraordinary and ordinary refractive indices of the LC, and d is the LC cell thickness. For field-dependent measurements, electric fields of amplitude 16 $V_{PP}$/μm, 18 $V_{PP}$/μm and frequency 20 Hz was applied across the cell. For evaluating the orientational order parameter (*S*) values, the experimental birefringence (Δ*n*) data was fitted with the well-known Hallers equation [67-69],

$$\Delta\boldsymbol{n} = \Delta\boldsymbol{n_0}(\boldsymbol{1} - \boldsymbol{T}/\boldsymbol{T^*})^{\boldsymbol{\beta}} \tag{3}$$

Here, $\Delta n$ is the birefringence, $\Delta n_0$, $T^*$ and $\beta$ are empirical fit parameters, $T$ is the absolute temperature. Physically, $\Delta n_0$ is the extrapolated value of birefringence at absolute zero temperature. The orientational order parameter ($S$) is then evaluated from, $\boldsymbol{S = \Delta n/\Delta n_0 = (1 - T/T^*)^{\beta}}$.For analyzing flexoelectric effects, the rubbing direction ($\vec{\boldsymbol{r}}$) was made parallel to the analyzer pass-axis. An external square-wave electric field of amplitude 18 $V_{PP}$/μm and frequency 5 Hz was applied during these measurements.

## 3. Results and discussion

### *3.1. Optical textures*

The characteristic LC textures of CB7CB, recorded between crossed polarizers are shown in Figure 2. Sharp changes in colour were observed close to the isotropic-nematic (Iso-N) transition due to an increase in the birefringence. In the nematic (N) phase, a uniform marble-like texture appeared with a gradual, temperature-dependent change in the birefringent colour. These changes in colour indicate a significant change in birefringence and that highly ordered microstructures might be present in the N phase of the LC [3,70]. The transition of the N phase to $N_X$ phase started near 103 °C. A polygonal, rope-like, pseudo-focal-conic texture appeared below 103 °C, typical of the $N_X$ phase [71-75]. Transition to a crystalline state was realized after slowly cooling the sample for a sufficiently long time. It was noted that after a few months, the $N_X$ phase could be supercooled to room temperature *via.* slow cooling.

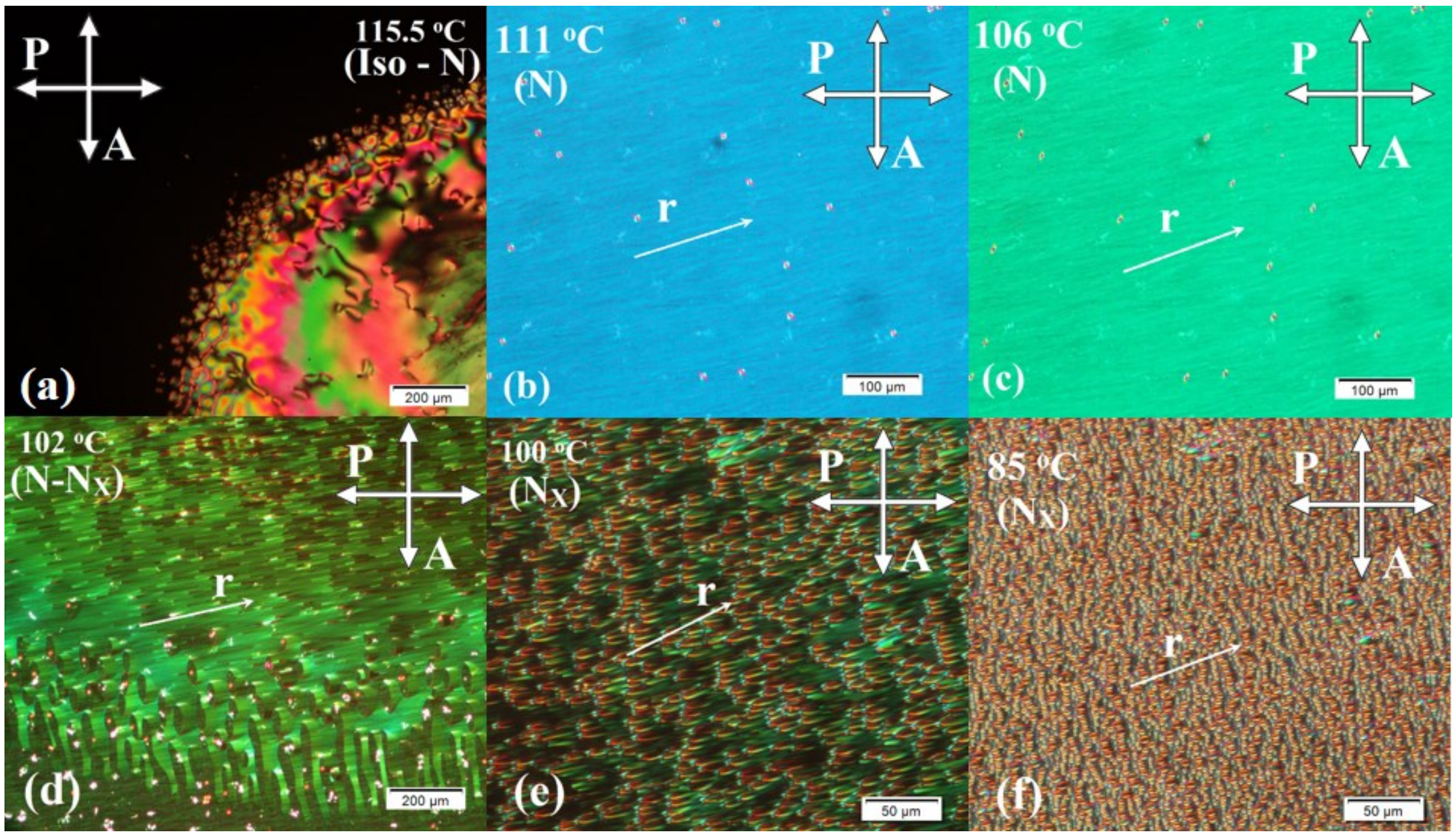

Figure 2. The optical textures of different LC mesophases exhibited by CB7CB, recorded between crossed polarizers, using a POM at different temperatures. Figure 2a was captured with the LC sandwiched between an untreated glass slide and a coverslip; Figures 2b-2f were recorded in a 5μm cell with planar alignment.

### *3.2. Electro-optic measurements*

Electro-optical measurements were carried out to investigate polarization behaviour of the LC compound. Planar LC cells of two different cell thicknesses 5μm and 3.2μm were used for the study. When a triangular input electric field of amplitude 12 $V_{PP}$/μm or higher was applied, a single delayed peak per half-cycle of the input field appeared (Figure 3a). The peak can be clearly identified superimposed on the linear ohmic background [45-55,76]. It was not centred about the zero-crossing of the applied field and persisted till frequencies up to ~ 150 Hz. The height and the time-scale position of the peak varied significantly with temperature. Such behaviour implies a long-lived ferroelectric state [47-49,54,55,77,78]. The height of the current peak increased with increasing temperature and survived in the isotropic phase. It is similar to ferroelectric-like behaviour reported in a wide number of BLCs [44-47,50,51,54-56, 76]. The associated polarization ($P$) values were calculated from the obtained current response and its variation with temperature is shown in Figure 3b. $P$ increased with rising temperature, similar to several other BLCs [44,46,49-51,76]. Also, $P$ decreased with increasing frequency and increased with increasing field strength [46,48,49,54]. At a fixed temperature, with increasing frequency, the polarization peak shifted to the right towards higher values in the time scale, and with decreasing peak height (Figure 4). It suggests that at higher frequencies the polar switching could not follow the input triangular field completely and as a result the value of $P$ decreased [47].

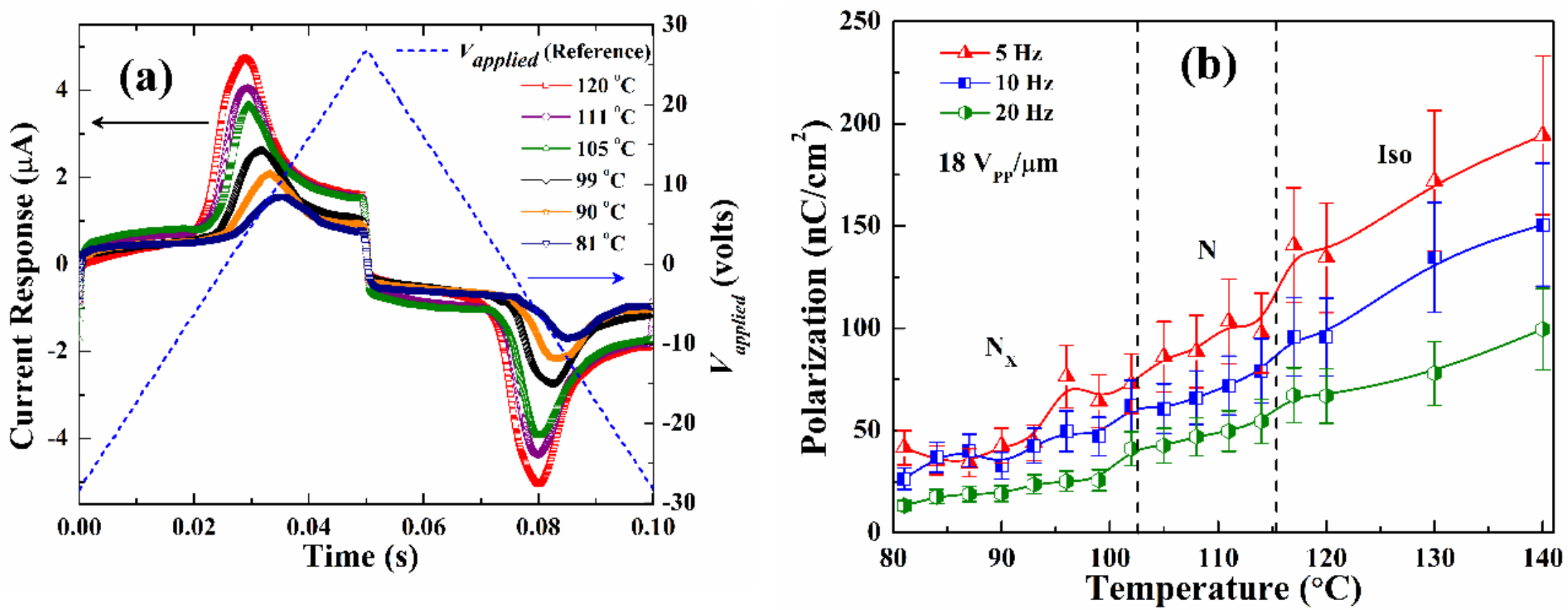

Figure 3. (a) Time-dependent switching current responses of the LC, at different temperatures (triangular input 18 $V_{PP}$/μm 10 Hz); (b) Temperature dependent variation of the polarization values calculated at different frequencies (input triangular electric field 18 $V_{PP}$/μm).

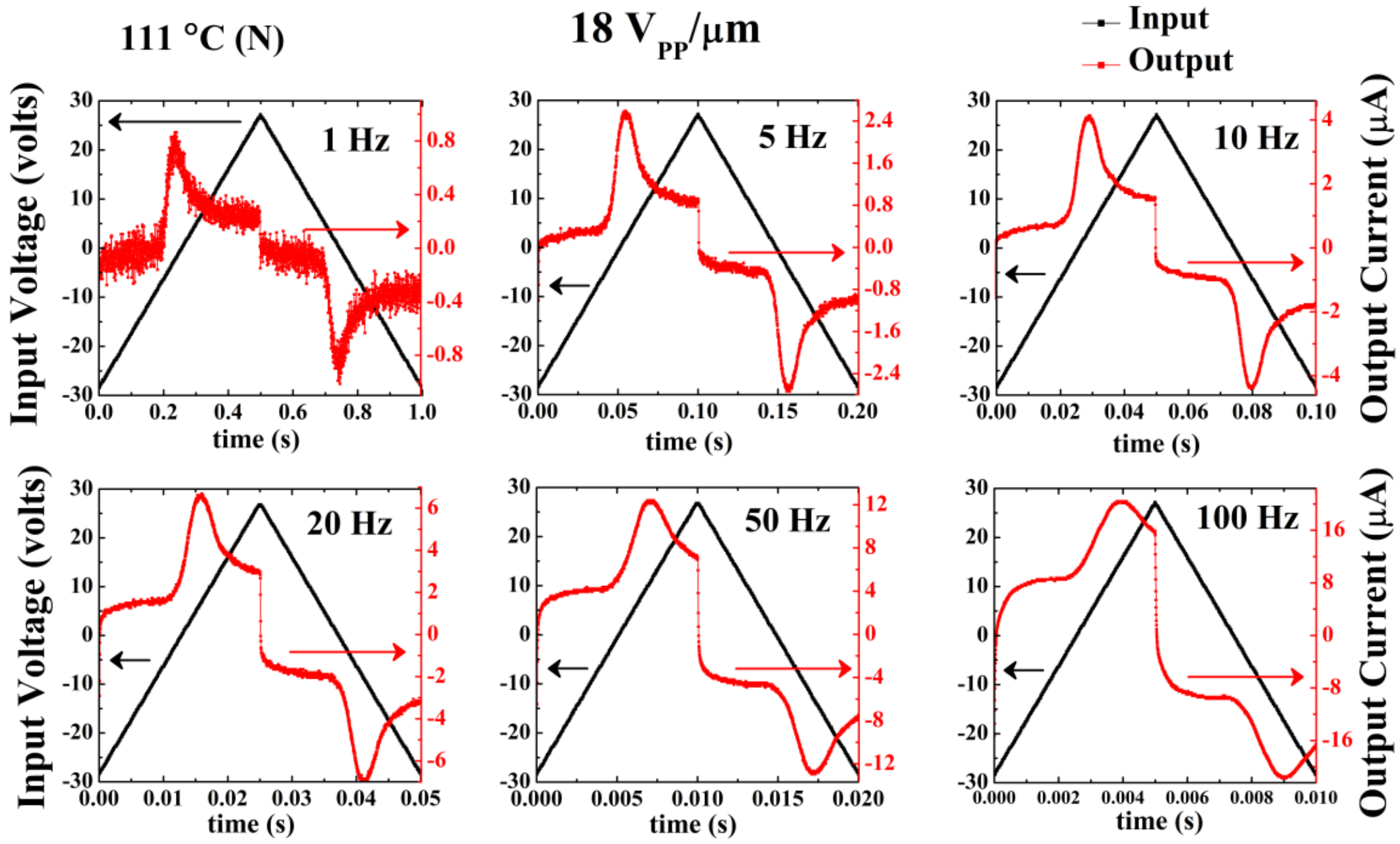

Figure 4. Frequency dependent variation of the polarization peak at 111 °C (N) with an input triangular electric field $18V_{PP}/\mu m$.

The LC sample was also probed with a square-wave electric field of amplitudes up to $18V_{PP}/\mu m$ and frequencies till ~ 150 Hz. Similar to the triangular-wave measurements, a delayed, distinct polarization bump was obtained in the current response (Figure 5a), appearing in each half-cycle. The bump appeared within 1 ms of field reversal, especially in the N phase, due to switching of the bulk polarization (*P*) [56,79]. The polarization switching times ($\tau_P$), evaluated from the time scale position of the polarization bumps, are shown in Figure 5b. The switching time ($\tau_P$) is in sub-millisecond range in the N phase and it remains almost fixed (~ 500 μs). With decreasing temperature, $\tau_P$ showed an increase (in the $N_X$ phase) due to increasing viscosity [44,56]. In the N phase, when a square-wave electric field was applied, the brightness of the texture decreased, and a frequency following switching was observed (Supporting Information Figure S13 and S14/Video1). On field removal, the usual nematic texture reappeared. Under the same conditions, the response obtained in the $N_X$ phase was different. The heliconical structure was deformed (or unwound) on field-application, and a uniform dark texture was established (Supporting Information Figure S15 / Video2). Settling of a homeotropic texture under ferroelectric switching conditions agrees with several other LCs reported in the literature [46,47,54]. The usual $N_X$ texture (usually identified as the twist-bend nematic ($N_{tb}$) phase in literature [11,12]) reappeared only very slowly once the field was turned off due to slow reorganization of the helical structure [35]. Recently, in a theoretical investigation Pająk *et al.* have predicted that the twist-bend helices can unwind under a strong electric field, giving rise to splay-bend nematic ($N_{SB}$) and polar nematic ($N_P$) phases [24]. It was followed by Merkel *et al.*, where they report experimental observations of a field-induced splay-bend nematic ($N_{SB}$) phase [69] and a large polar order [35] in the $N_{tb}$ phase. This is analogous to field-induced deformation and consequent unwinding of helical structure in FLCs [33]. Recently, a few other theoretical models have also predicted that a strong electric field, perpendicular to the helix axis, is expected to completely (or perhaps only partially) unwind, the helical structure of the $N_{tb}$ phase [12,24,25,30,35]. Again, the $N_{tb}$ phase is expected to be locally polar [12-14,24,29], and our experimentally obtained threshold field values (> $12V_{PP}/\mu m$) for the polarization response are of the same order predicted in [24]. Therefore, unwinding of the helical structure is expected to give rise to a polar response, as obtained in our experiments.

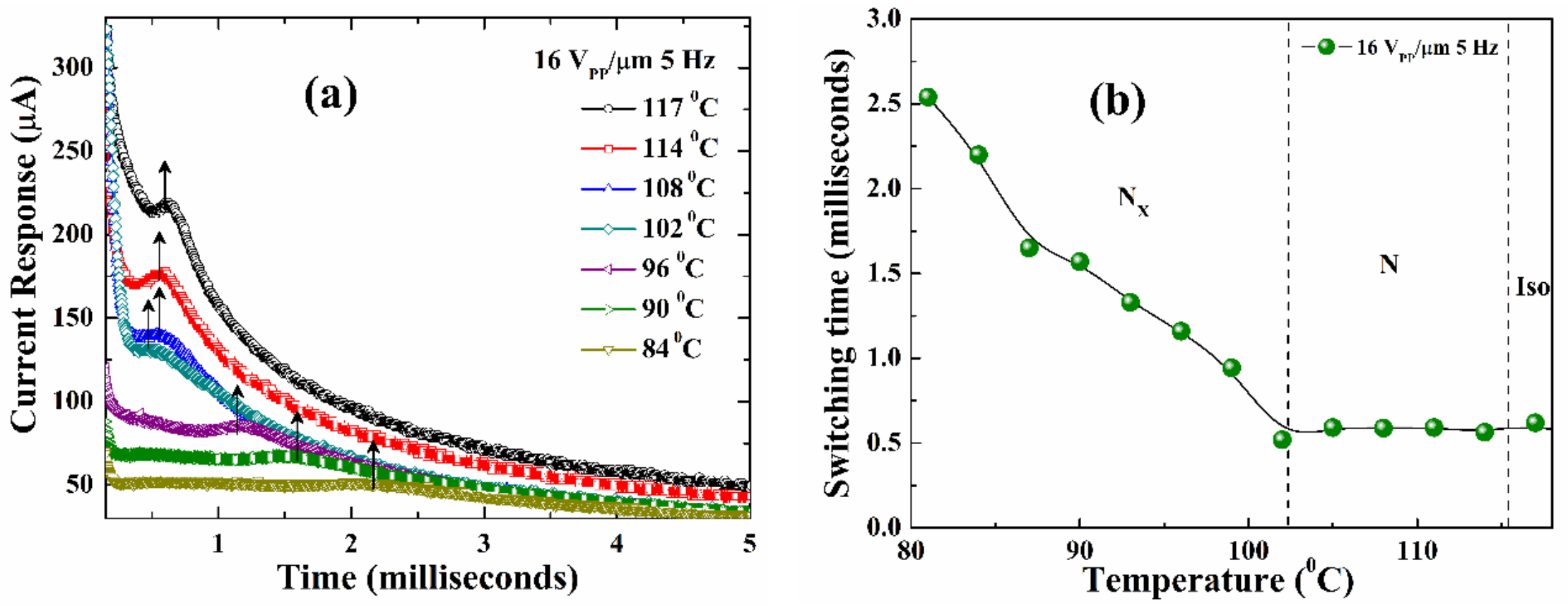

Figure 5. (a) The current response and (b) Polarization switching time ($\tau_P$) of the compound under a square-wave electric field of frequency 5Hz and amplitude 16 $V_{PP}$/μm, at different temperatures. In 5(a), the arrows indicate position of the polarization bumps.

The polar response is persistent in the high-temperature nematic (N) phase of the sample and its origin needs to be understood. In a recent study, Heist *et al.* have described the nature of this high-temperature N phase in CB7CB [41]. The authors discuss that this high-temperature N is a cybotactic nematic phase consisting of polar-twisted nematic ($N_{PT}$) clusters ($N_{CybPT}$). They arrive at this conclusion based on several different experimental observations and indicate that these clusters have the same structure as the lower-temperature $N_{PT}$ domains, but with a much smaller domain-size [41]. These observations are in coherence with the findings of clusters in the N phase of CB7CB by Krishnamurthy *et al.* [43]. The compound CB7CB is a bent LC dimer and in literature BLCs are known to form small aggregates/clusters (cybotactic clusters) with a higher internal symmetry (e.g. smectics) in their N phase [44,46,48-55,76]. These clusters are formed due to the bent molecular shape and the steric barrier produced by surrounding molecules [53-55,80-84]. The cybotactic clusters are locally polar because of their layered internal structure and collective orientation of the dipoles inside each cluster [46,48-50,52,54]. In the absence of an electric field, the clusters are randomly oriented in the LC bulk. When an external electric field above a certain threshold is applied, they align collectively giving rise to a net polar response [1,3,44-56]. The clusters can be present in the nematic as well as in the isotropic phases [44,46-55,76]. Their existence in the N and the isotropic phases have been reported using various techniques such as the small angle X-ray scattering (SAXS) [80], second harmonic generation (SHG) [54,55], transmission electron microscopy (TEM) [81], dynamic light scattering (DLS) [82], dielectric spectroscopy [44], spontaneous polarization measurements [54] and nuclear magnetic resonance (NMR) [81]. Therefore, a net polar response in the $N_X$, N and in the isotropic phases can be attributed to the field-induced alignment of cybotactic clusters. In this paper, we have explored the presence of these clusters in CB7CB using dielectric techniques which will be discussed later in the next sub-section. However, in the $N_X$ phase, based on our experimental results and the findings reported in literature [24,35,41,43,69], we propose that the polarization has two possible origins: (i) locally polar (cybotactic) clusters persisting in the $N_X$ phase and (ii) deformation and consequent unwinding of the helical structure under a strong external electric field (> 12 $V_{PP}/\mu m$). When subjected to a strong external electric field, collective orientation of clusters and unwinding of the helices take place simultaneously. As a result, a net polarization with appreciable modulus is obtained in the $N_X$ phase. With increasing temperature, cluster sizes and the viscosity of the medium decrease, making the field-induced alignment of clusters easier - as a result the polarization value increases [1,44-46,48-51,56,76]. Similarly, when temperature is lowered, size of the clusters and the viscosity increase. Thus, the polarization value becomes smaller at lower temperatures because now the clusters cannot reorient easily with the applied electric field [1,44-46,48,49,56]. Also, with increasing field strength, the value of polarization increases (comparing Figure 3b with Supporting Information Figure S11-b) [49]. With an increase in the field strength, the number of clusters participating in the switching mechanism increases. The same is true for the number of helices being unwound and contributing to the net polarization. Therefore, at a higher field strength, more number of clusters and unwound helices now contribute to the net polarization rendering large $P$ values.

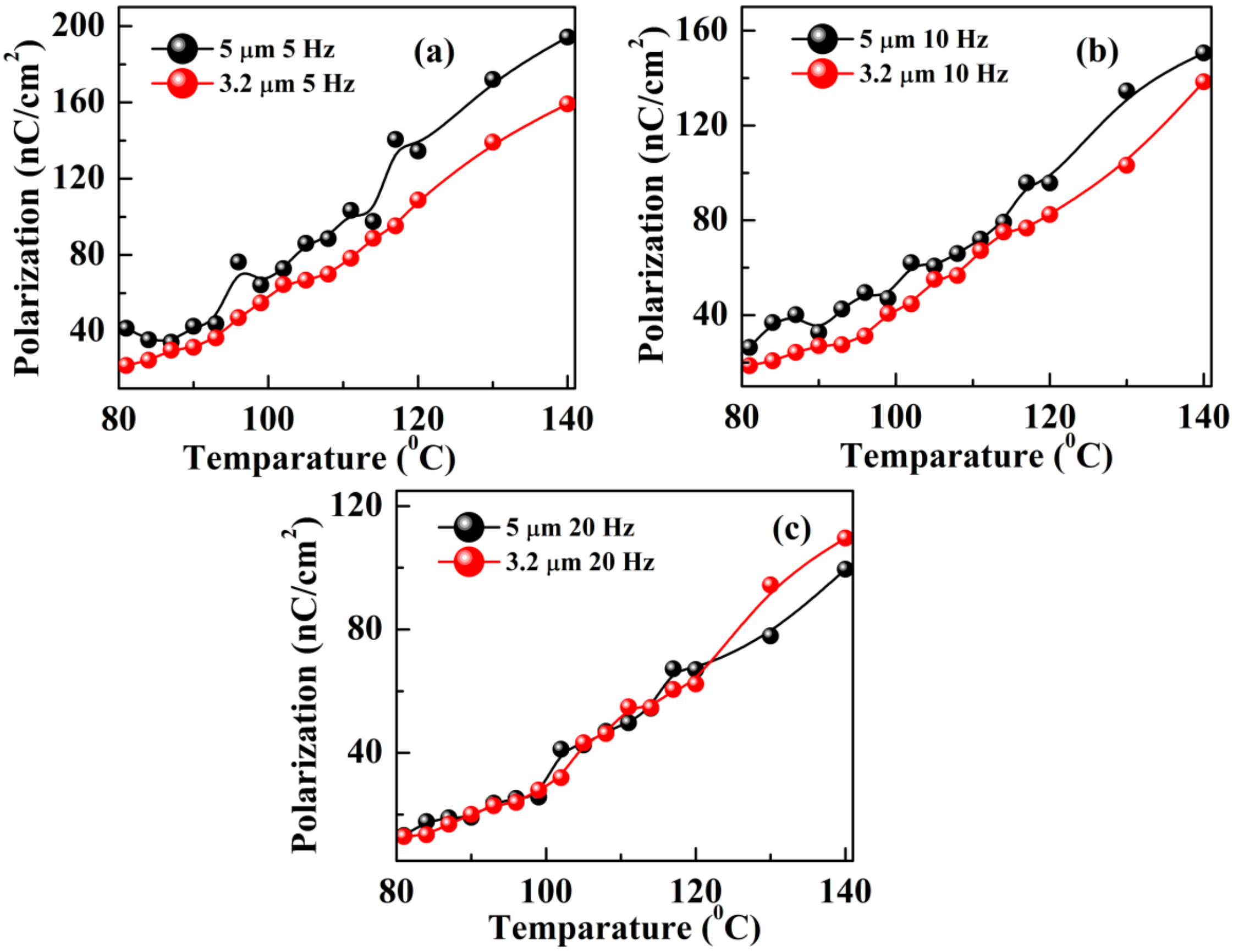

Figure 6. Temperature dependent variation of the polarization values using cells of different thickness – 3.2 μm and 5 μm, at various frequencies: (a) 5 Hz, (b) 10Hz, and (c) 20Hz. At 20 Hz, the polarization profiles overlap.

The results till now indicate that the polar response has a ferroelectric-like origin but contributions from ionic impurities are also likely to be present. Separation of polar and ionic contributions is difficult. It is possible that the ionic peak merged with the polarization peak preventing their differentiation [44,46,47]. We have investigated this possibility *via* experiments and results reported in the literature. For purely ionic cases, the polarization response appears only in the low-frequency regime and lacks in a strong temperature dependence [47-49]. Also, ionic impurities in the LC phase cannot move at frequencies more than 10 Hz [46,85-87]. Our results, in contrast, show a strong temperature dependence of the polarization and it persists at moderately high frequencies (up to ~ 150 Hz) (Figure 4). Electroconvection (EC) patterns were not observed during the experiments in the entire frequency regime. This indicates that the ionic impurities are very less and they hardly have any effect on the observed behaviour [49,53]. The polarization values obtained in our experiments (~ 200 nC/cm$^2$) are much larger than the typical polarization values for ionic and surface contributions (~20 nC/cm$^2$) [78]. If the polar contributions arise solely due to ions, the net polarization values would be proportional to the cell thickness [53,54,56]. Therefore, to investigate the extent of ionic contributions, we have used LC cells of two different thickness – 5 μm and 3.2 μm. We found that with increasing frequency polarization values decreased, and eventually, at around 20 Hz the polarization profiles for both the cells merged together (Figure 6). Also, the *P* values decreased with increasing frequency, similar to several other BLCs [46,48,49,54]. It happened because at high frequencies the ionic contribution to net polarization goes down and only the contribution from a ferroelectric-like polar order remains. AC conductivity ($\sigma_{ac}$) measurements were also performed in the experimental frequency region (up to ~ 100 Hz) (Figure 7). $\sigma_{ac}$ did not change significantly with temperature, which again shows the ionic contributions to the net polarization, if any, is only finite. With increasing frequency, $\sigma_{ac}$ increased which is opposite to the net polarization going down. Interestingly, in the isotropic phase $\sigma_{ac}$ decreased which is in contrast with the net polarization going up (Figure 7 inset). The values of $\sigma_{ac}$ in our experimental frequency regime are nearly one order less than what is reported in the literature for CB7CB [88]. Further, the dc conductivity ($\sigma_{dc}$) value, obtained from dielectric fittings (discussed in the next sub-section), were found around ~$10^{-9}$ S/m (matches with [69]) which is two orders of magnitude smaller than $\sigma_{ac}$. This is a strong indication that ionic impurities are not the principal reason behind the observed polar response. Also, the frequency range of existence of the polarization peak matches almost exactly with the range of collective mode ($M_1$) identified in the dielectric spectroscopy measurements (discussed later). Therefore, based on these experimental findings, we conjecture that the net polarization has a ferroelectric-like origin (polar cybotactic clusters and additionally, unwound helices in the $N_X$ phase) and it may have limited contributions from ionic charges as well [46,50]. There is a competition between the polar and the ionic contributions in the entire LC phase, which becomes dominantly polar at higher frequencies.

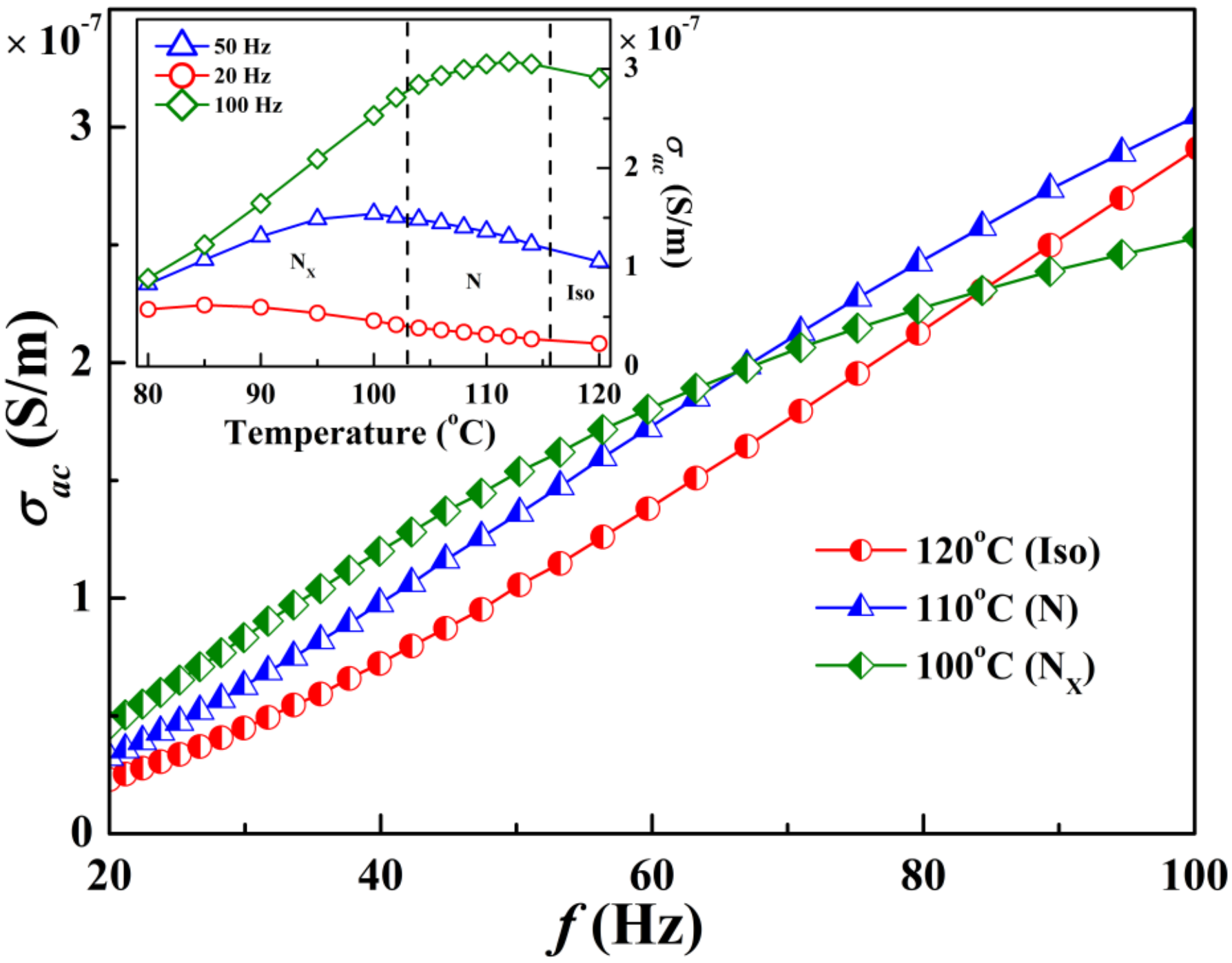

Figure 7. Frequency dependent variation of the ac conductivity ($\sigma_{ac}$) at different temperatures; the inset shows temperature dependent variation of $\sigma_{ac}$ at different frequencies.

### *3.3. Dielectric measurements*

Broadband dielectric spectroscopy was performed in the frequency range 20 Hz – 2 MHz. The real and the imaginary ($\varepsilon'$ and $\varepsilon''$) parts of the complex dielectric permittivity were measured at different temperatures (Supporting Information Figure S6a and S6b). At smaller frequencies the value of $\varepsilon'$ is very large (~ 100), typical of ferroelectric substances [9,48,49]. The dielectric absorption spectra ($\varepsilon''$) revealed two distinct relaxation modes: a low-frequency mode ($M_1$) and a high-frequency mode ($M_2$). $M_1$ suggests collective relaxations, while $M_2$ represents reorientation of the LC molecular short-axis due to planar anchoring conditions [9,44,46,48-51,76]. $M_1$ varied significantly with temperature while $M_2$ exhibited a minute temperature dependence. The dissipation factor $\tan\delta$ ($= \varepsilon''/\varepsilon'$), which is another measure of dielectric loss, was also analyzed under an external DC bias voltage (Supporting Information Figure S7). The low-frequency $\tan\delta$ loss-peak (representing $M_1$) was suppressed continuously under the DC bias voltage (up to 20 V). This is a confirmatory experimental proof that $M_1$ arises because of collective relaxations [9,44,46,51]. Also, with DC bias application, the relaxation frequency was reduced, similar to the lowest frequency (~ 100 Hz) collective mode reported in [35]. Collective relaxations in the N phase of BLCs appear due to the polar cybotactic clusters [46,48-55,76]. These clusters can exist in the isotropic, nematic (N) and in the $N_X$ phase of LCs [41,43,80-84]. In fact, a few recent papers have already discussed the presence of these clusters in the LC phases of CB7CB and also the cybotactic nature of the associated high-temperature N phase [41,43]. In the $N_X$ phase, the clusters reside with the spontaneously formed helices and give rise to properties similar to the N phase (e.g. polarization, collective relaxations). Recently, low-frequency collective relaxation modes, similar to $M_1$ (up to a few hundred Hz), have been reported in the $N_{tb}$ phase of BLCs, bent LC dimer CB7CB and its higher homologues [35,69,89]. They were attributed to the collective fluctuations of azimuthal angle, had a large associated polar order (dielectric strength, $\delta\varepsilon$ as high as ~ 800) and have been discussed as reminiscent of the Goldstone mode manifested by ferroelectric LCs (FLCs) [35,69,89]. The collective modes in the $N_{tb}$ phase (or $N_X$ phase) is largely dependent on the cell thickness. In literature, it has been reported for relatively thin LC cells only (~ 5 μm) [35,69,89], and it is absent in cells with larger electrode separation/thickness [27,90-92]. Also, the relaxation frequency of the collective mode decreases with increasing cell thickness [69]. As a consequence, cumulating the results, we propose that collective relaxations in $N_X$ phase has two possible origins: (i) the polar cybotactic clusters persisting in the $N_X$ phase [41,43] and (ii) fluctuations in the tilt or the helicoidal angle of the twist-bend helix [35,69,89]. These predictions are also coherent with the origin of polarization in $N_X$ phase discussed in section 3.2. The mode $M_2$ was not suppressed under DC bias and hence it was attributed to the reorientation of the LC molecular short-axis [9,44,46,48-51,76]. The temperature dependence of $f_{R2}$ and $\delta\varepsilon_2$ (Supporting Information Figure S9-b) also indicate that the origin of $M_2$ is reorientation of the molecular short-axis. However, it is also possible that $M_2$ resulted from the ITO electrodes because of its finite resistance, and also because ITO relaxations may become temperature dependent in a cell filled with LC.

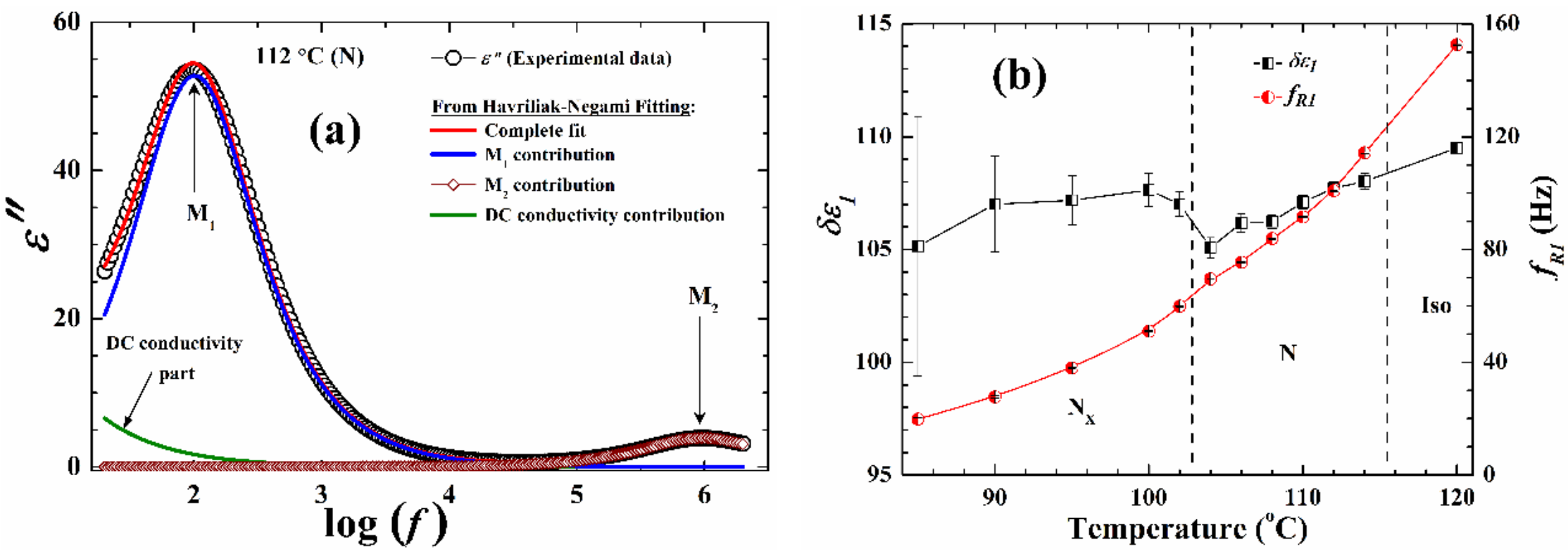

Figure 8. (a) Dielectric loss ($\varepsilon''$) as a function of frequency in the nematic (N) phase, and (b) temperature dependence of the dielectric strength ($\delta\varepsilon_1$) and relaxation frequency ($f_{R1}$) of mode $M_1$. In 8a, the blue, brown and green curves represent the contributions from $M_1$, $M_2$, and the dc conductivity, respectively, extracted from the total H-N fitting (red); open circles represent the experimental data.

The dielectric loss data was fitted to the well-known Havriliak-Negami (H-N) equations for evaluating the associated dielectric parameters, as described in section 2.4. The overall dielectric loss has contributions from the two relaxation modes and a dc conductivity part. The three individual contributions extracted from the complete fit are shown in Figure 8a. The parameters $\alpha$ and $\beta$ (for both $M_1$ and $M_2$) vary between 0.98 – 0.99 and thus represent a near Debye-type relaxation. In the low-frequency regime (i.e. $M_1$), tilted smectic-like ferroelectric clusters are known to give rise to such Debye-type relaxations [93-95]. Similar analogy for the low-frequency collective modes were drawn in reference [69]. The parameters dielectric strength ($\delta\varepsilon$) and relaxation frequency ($f_R$) of $M_1$, extracted from the fit of equation (1), are shown in Figure 8b. The range of $f_{R1}$ matches well with the range of existence of polarization $P$ (up to ~ 150 Hz). The dielectric strength of $M_1$ is large ($\delta\varepsilon_1$ ~ 105-110) and remains fairly constant with varying temperature, except for a small discontinuity near the N-$N_X$ transition [48,49]. Such high values of $\delta\varepsilon_1$ indicates an associated polar order [35]. On the other hand, the relaxation frequency ($f_{R1}$) decreases upon cooling, divulging an Arrhenius-type behaviour ($\boldsymbol{f_R = f_0 \exp(-E_a/k_BT)}$; $\boldsymbol{f_0}$ is a temperature-independent constant, $\boldsymbol{E_a}$ is the activation energy, $\boldsymbol{k_B}$ is the Boltzmann's constant and $T$ is the absolute temperature). The activation energy ($E_a$) of $M_1$ was evaluated from the slope of the linear fit of the Arrhenius plot, as shown in Figure 9a. In the $N_X$ phase, $E_a$ has a value ~ 71.34 ± 1.24 kJ/mol which is nearly twice the value reported for mode $P_1$ (~ $10^5$ – $10^6$ Hz) in reference [91]. In the N phase, $E_a$ (~ 59.56 kJ/mol) has values smaller than the $N_X$ phase (possibly due to lesser viscosity at high temperatures), but it is still quite large. Such high activation energy dictates that a large amount of energy is required to excite the collective mode ($M_1$). This is why it is rather hard to realize $M_1$ in a relatively thicker cells. As already discussed in this section, the collective mode in the $N_X$ phase is largely dependent on the cell thickness and it has been reported only for relatively thin LC cells only [35,69,89]. It also explains the large threshold voltage required for a polar response, as found in our experiments. In case of $M_2$, the dielectric strength ($\delta\varepsilon_2$) is very small, and decreases continuously on cooling (Supporting Information Figure S9-b). The dc conductivity ($\sigma_{dc}$) extracted from the fitting data, was found around ~$10^{-9}$ S/m, which is of the same order (~ 1 nS/m) reported in [69]. The extent of ionic impurities present in the LC were also examined by evaluating the concentration of free ions ($n$) and the diffusion coefficient ($D$). The $n$ and $D$ values were calculated by fitting the modified Uemura equation [96-98] to the conductivity contribution (of $\varepsilon''$) extracted from the H-N fitting (Equation 1). Both $n$ and $D$ decreased with decreasing temperature remaining nearly constant in the N phase (Figure 9b). $n$ lies in the range of ~ 0.8 – 1.5 × $10^{19}$ $m^{-3}$ which is two order less in magnitude than what is reported in a majority of LCs [97-102]. Also, $n$ decreases in the isotropic phase, which agrees with the decrease of $\sigma_{ac}$ on entering the isotropic phase (Figure 7 inset). Also, dielectric permittivity measurements (of $\varepsilon_{||}$ and $\varepsilon_{\perp}$) at 1 kHz were performed to compare our results with the ones reported first by Cestari *et al.* (Supporting Information Figure S10) [27]. The dielectric anisotropy ($\Delta\varepsilon$) values, as a function of temperature, were also evaluated (Supporting Information Figure S10-inset). Our results are in very good agreement, both quantitatively and qualitatively, with the ones reported in the literature [27,62]. These agreements particularly show the high purity of our sample under investigation and discards the possibility of impurities being present, if any. Therefore, it is established that the contribution from free ions to the net polarization response is rather small and finite, and it decreases appreciably with decreasing temperature and increasing frequency. It further establishes that the polar response observed in our experiments indeed has a ferroelectric origin, similar to experiments reported in a wide number of bent-core nematic LCs [44,46-56,76]. Also, our results provide experimental grounds to a large number of theoretical models and predictions of a polar response, especially in the Nx phase, such as reference [24,37-39,41].

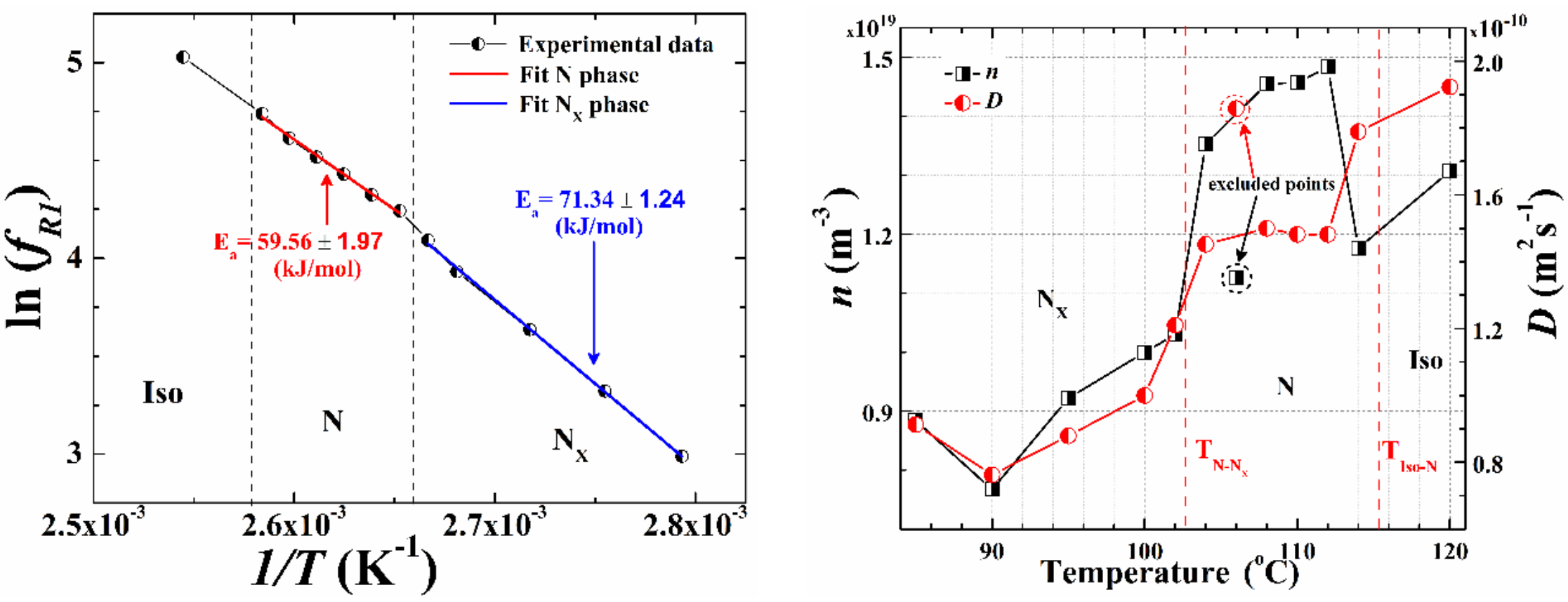

Figure 9. (a) Arrhenius plot of the $M_1$ relaxation frequency ($f_{R1}$). $E_a$ is calculated from the slope of the linear fit to the data represented by solid lines; (b) Temperature dependence of ion concentration ($n$) and diffusion coefficient ($D$) obtained from Uemura fitting of the dc conductivity part (shown in Figure 8a).

### *3.4. Optical transmission studies*

The optical birefringence ($\Delta n$) measurements of the LC CB7CB was performed by analyzing the temperature dependent optical transmission data, as described in section 2.5 [66]. The obtained temperature dependent variation of $\Delta n$, on cooling the LC from isotropic liquid, is presented in Figure 10(a) (open circles). In the N phase, $\Delta n$ increased with decreasing temperature, manifesting the usual nematic behaviour [62-64], until the N-NX transition. At the N-NX transition, $\Delta n$ increased slightly (~ 0.003), followed by a continuous decrease with decreasing temperature in the NX phase [63,64]. The birefringence values were well fitted with the classical Haller formula (equation 3) in the N phase [63], and the associated fitting parameters were determined. We found $\Delta n_0$ = 0.2659, $T^*$ = 114.9443 $^o$C and $\beta$ = 0.17499, that agree well with the values reported in literature (Supporting Information Figure S16) [63,69,89]. Figure S16 also demonstrates a comparative plot between our results and the results reported in literature [62,63]. In the N phase, the order parameter ($S$) values were determined using the relation, $\boldsymbol{S} = \Delta\boldsymbol{n}/\Delta\boldsymbol{n_0} = (\boldsymbol{1} - \boldsymbol{T}/\boldsymbol{T}^*)^{\boldsymbol{\beta}}$. The calculated $S$ values are shown in Figure 10(a) (red continuous line) and it was also extrapolated to the NX phase (blue dashed line) under the assumption that $S$ is continuous at the transition [63]. In reality, in the NX phase, however, the order parameter does not follow the usual nematic trend, as it is evident from the $\Delta n$ variation. It is essential to understand how $\Delta n$ and $S$ behaves under an external electric field for drawing analogy with the already presented experimental results. For the electric field dependent measurements, the LC sample was cooled down from isotropic liquid under the influence of an externally applied square-wave electric field. The obtained results are presented in Figure 10(b). The order parameter values for this case were calculated using the relation, $\boldsymbol{S} = \Delta\boldsymbol{n}/\Delta\boldsymbol{n_0}$, where we used $\Delta n_0$ = 0.2659 from the results in Figure 10(a). Near the isotropic-N transition, the birefringence value increased sharply and thereafter decreased continuously with decreasing temperature. The initial increase in the field-induced birefringence is due to pre-transitional effects, similar to other nematic LCs [103,104]. As already discussed in section 3.2, in the N phase, a square-wave electric field results in a grainy texture with decreased intensity (Supporting Information Figure S13 + S14 / Video1). This observation is reflected in the measurements presented in Figure 10(b) – the birefringence decreases in the N phase under an external electric field. It is possibly because of an out-of-plane rotation of the optic axis under the influence of an external field. On entering the $N_X$ phase, the birefringence value dropped quickly to very small values. It is in coherence with the settling of a uniform dark texture in the $N_X$ phase when an external electric field is applied (already discussed in section 3.2). Therefore, the field-induced optical transmission measurements concur with the unwinding of helical structures when subject to a strong external electric field (video 2). Further, a few recent studies have shown that an electric field applied perpendicular to the helix axis can unwind the helical structure of $N_X$ phase and induce a polar nematic ($N_P$) phase [24,35]. Therefore, we conjecture that an external electric field unwinds the helices in the $N_X$ phase and an out-of-plane rotation of the optic axis takes place by an angle 90°. It resulted in the uniform dark (homeotropic) texture to settle in the $N_X$ phase. On removal of the applied field, it was observed that the usual $N_X$ texture reappeared very slowly due to slow reorganization of the helical pseudolayered structure [35]. It was also observed that the N-$N_X$ transition temperature was slightly lowered (~ 0.8 $^{o}$C @ 16 V/μm) because of the applied field (Figure 10(b) inset). It is possibly due to field-induced strong anchoring which results in a lowering of the transition temperature.

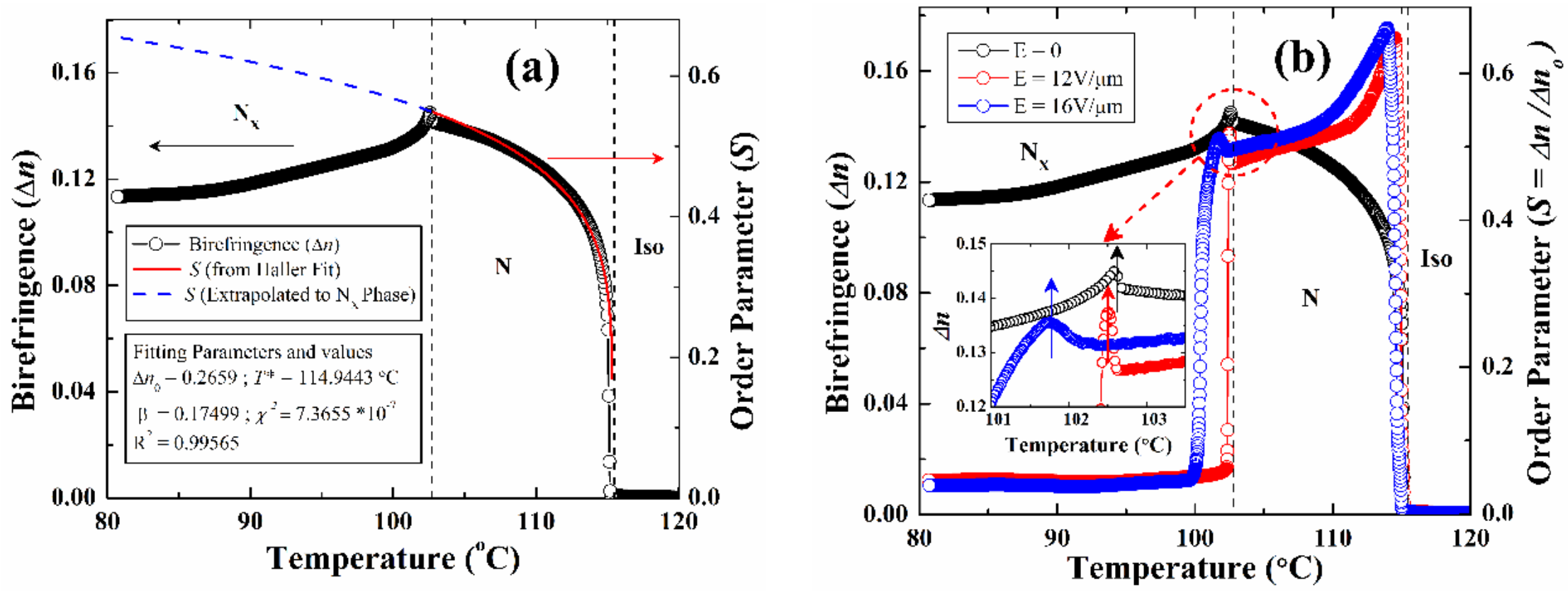

Figure 10. (a) Temperature dependent variation of birefringence ($\Delta n$) and order parameter ($S$) of CB7CB; (b) birefringence ($\Delta n$) and order parameter ($S$) variation of CB7CB as a function of temperature at different applied electric field values (black: E = 0, red: E = 12 V/μm and blue: E = 16 V/μm; @ f = 20 Hz); the inset of (b) shows the variation near N-$N_X$ transition, arrows indicate the transition temperatures.

In an external electric field, additional effects may arise, such as deformations (or unwinding) of the twist-bend helical structure, flexoelectric effect and electroclinic effect [24,35,105]. To investigate if any such effect actually arises, electric field dependent transmission measurements were performed, as discussed in section 2.5, such that the rubbing direction ($\vec{r}$) was parallel to the analyzer pass-axis (it makes the setup sensitive to in-plane rotations of the optic axis). In the $N_X$ phase, the E-on intensity manifested a net decrease with respect to the E-off intensity (Supplementary Figure S17-b). It agrees well with the POM observation of a uniform dark texture in the E-on state in $N_X$ phase, as already discussed in this section (Supporting Information Figure S15 and video 2). The mismatch between the E-off state intensities in the $N_X$ phase is caused by the slow reorganization of the pseudolayered structure. A closer examination of the E-on intensity reveals a field-following flickering at the output at low frequencies (Supporting Information Figure S17a and S17b inset + video 3). Flickering arises with fluctuating transmittance when the field's polarity is altered regularly [106-108]. Although several factors may give rise to flickering in LCs, the dominant factor is the flexoelectric effect (FEE) [106-109]. A mechanical distortion in the LC director caused by splay or bend deformation results in a net polarization, called the flexoelectric polarization [106-109]. FEE is the interaction of this net polarization with the external electric field, and it is dependent on the applied field's polarity. FEE has been widely encountered in bent-core LCs [25,110,111], and in rod-like nematic LCs [106-108] which causes flickering in the transmitted light intensity. CB7CB is known to have a small bend elastic constant ($K_{33}$) [26,28] and hence it is prone towards bend deformations. The $N_X$ phase has a helical internal structure and additionally, the molecules are bent-shaped and have inherent molecular tilt. Therefore, any small but finite incompatibility with the planar boundary may create splayed and bent regions near the boundaries. This, in turn, can cause flexoelectric polarization to appear [105]. In fact, flexoelectric polarization in the $N_X$ phase of CB7CB has been demonstrated very recently [25,35]. Therefore, the observed flickering, although undesirable in display applications, helps identify flexoelectric effects arising in the $N_X$ phase. The flickering decreased with increasing frequency, and at frequencies ~20 Hz and above no flickering could be detected (video 4). Also, at 20 Hz, the temperature dependent polarization profiles of the two cells (3.2 μm and 5 μm) merged together (Figure 6-c). The LC response time was found to lie in the range of 80-100 μs in the entire LC phase when subjected to a 18 $V_{PP}$/μm, 1 kHz square-wave electric field (Supporting Information Figure S18). It has strong resemblance with the ferroelectric LCs having sub-millisecond switching times [112,113] and it also complies with the polarization response obtained in our experiments. The $N_X$ phase of CB7CB was initially considered to be biaxial in a few earlier communications [114,115]. The molecular biaxiality of CBnCB dimers (n = 5, 7, 9 etc.) increases with decreasing spacer length and it also affects the helical pitch of the $N_X$ phase [116]. However, in our studies we could not detect any signature of biaxiality which could be partly due to flexibility of the heptane spacers. No further experiments were performed in this regard.

## 4. Summary and Conclusion

A bent LC dimer CB7CB has been investigated in the N and $N_X$ phases using various techniques. In triangular-wave measurements, a single, delayed, polarization peak, not centred about the zero cross-over, was observed above a certain threshold electric field. The polarization peak manifested strong dependence on temperature and it persisted till very high frequencies (~ 150 Hz) where the ionic contributions, if any, are very less. If the polar contributions arose solely due to ions, the net polarization values would have been proportional to the cell thickness. On the contrary, we found in our experiments that polarization values for two different cell thicknesses merged together at around 20 Hz. It reveals a true ferroelectric nature of the obtained polar response. We do not completely discard the contributions arising from ionic impurities. The free ion concentration ($n$), $\sigma_{ac}$ and $\sigma_{dc}$ have small values in the mesophase range. Also, frequency and temperature dependent variation of $\sigma_{ac}$ manifested contrasting behaviour with the net polarization ($P$). The polarization contributions from ionic impurities is therefore only finite and they become extinct at high frequencies. Similar characteristics were reflected in the square-wave measurements. The polarization switching times were found to lie in the sub-millisecond regime. The dielectric studies established collective relaxation processes in the entire LC phase, with large associated dielectric strength ($\delta\varepsilon$) values. Also, the range of frequency, for which polarization ($P$) exists (~ 150 Hz) in the electro-optic measurements, matches almost exactly with the frequency range of collective mode relaxations ($M_1$) identified *via.* dielectric spectroscopy. Based on the studies reported in literature, we discuss that these collective modes are only observed in relatively thin cells and have large activation energies [35,69,89]. The large activation energies associated with the collective mode explains the threshold voltages required for the polar response. The collective modes resulted due to the presence of cybotactic clusters, concurring with the results reported in literature [41,43]. Therefore, we attributed the net polar response to the field-induced reorientation of polar cybotactic clusters. Additionally, the helical structures were deformed under an external electric field. In the $N_X$ phase, field-induced deformations of the helical structures also contribute to the net polarization, which matches with the predictions in [24]. Therefore, when subjected to a strong external electric field, collective orientation of clusters and unwinding of the helices take place simultaneously giving rise to the net polar response. The results are well supported with experimental data and supplementary video evidences. The optical birefringence ($\Delta n$) and order parameter ($S$) measurements demonstrate excellent agreement with the results reported in literature. Field-dependent optical transmission measurements confirm deformation of the helical structure under an external electric field. Field-induced flickering observations revealed flexoelectric effects arising in the LC phase which extinct at frequencies ~ 20 Hz and above. At the same frequency value (20 Hz), the polarization profiles for the two LC cells with different thicknesses merged together. The evaluated response times also lie in the range of ~100 μs concurring with ferroelectric origin of the polar response.

In conclusion, we have demonstrated a net polar response in the N and $N_X$ phases of a bent LC dimer CB7CB. The polar response is ferroelectric-like in nature and it also has limited contribution from ionic charges. The field-induced deformation of twisted helical structures also contributed to the total polar response. On a high note, the study provides an experimental ground to several theoretical predictions of an underlying polar order in the $N_X$ (or $N_{tb}$) phase of LCs, such as references [24,37-39,41]. Further, it demonstrates various effects arising from the field-induced polarizations.

**Acknowledgements:** A. S. and S. P. gratefully acknowledges SERB, Department of Science and Technology, Government of India (EMR/2015/001897). S. P. acknowledges IITD for financial support, and Dr. Susanta Chakraborty for useful discussions on the fitting of experimental data.

# Supplementary Information

**Nematic twist-bend phase of a bent liquid crystal dimer: field-induced deformations of the helical structure, macroscopic polarization and fast switching speeds**

Sourav Patranabish[a], Aloka Sinha[a]*, Madhu B. Kanakala[b] and Channabasaveshwar V. Yelamaggad[b]

[a]*Department of Physics, Indian Institute of Technology Delhi, Hauz Khas, New Delhi 110016, India*
[b]*Centre for Nano and Soft Matter Sciences, P. B. No. 1329, Prof. U. R. Rao Road, Jalahalli, Bengaluru 560013, India*

* *Corresponding author: alokaphysics@gmail.com; aloka@physics.iitd.ac.in*

Supplementary Information

**Nematic twist-bend phase of a bent liquid crystal dimer: field-induced deformations of the helical structure, macroscopic polarization and fast switching speeds**

**General:** All starting chemicals procured from local and abroad sources were used as received. For monitoring reactions as well as to examine the purity of the compounds, thin layer chromatography (TLC) was used; aluminum TLC plates pre-coated with silica gel (Merck, Kieselgel60, F254) were used. For partial / complete purification, column chromatography technique was employed where silica gel (60–120, 100–200 mesh) was used as a stationary phase. Molecular structural characterization was carried out using various spectroscopic techniques. UV/Vis spectra were recorded with the help of PerkinElmer's Lambda 20 UV/Vis spectrometer (1 cm path length, $CH_2Cl_2$). The infrared spectra were recorded on KBr pellets in transmission mode (in the range of 400–4000 $cm^{-1}$) on a PerkinElmer Spectrum 1000 FT-IR spectrometer. Bruker AMX-400 (400 MHz) spectrometer was used to record $^{1}H$ and $^{13}C$NMR spectra in $CDCl_3$ at ambient temperature. The chemical shifts are reported in 'ppm' on scale downfield from TMS regarded as an internal standard. The coupling constants (*J*) are given in Hz.

**Synthesis:** The mesogen **CB7CB** was prepared by following synthetic steps depicted in the scheme below. Compounds 1-6 were synthesized by following protocols reported in the literature.[R1,R2]

**Scheme**

**Reagents and conditions:** (i) Oxalyl chloride, RT, 12 hrs; (ii) bromobenzene, $AlCl_3$, 0 °C, 12 hrs; (iii) hydrazine hydrate, EtOH, reflux, 12hrs; (iv) *t*-BuOK, toluene, reflux, 48 hrs; (v) n-BuLi, $B(OMe)_3$, -78 °C, 12hrs; (vi) $Pd(PPh)_3$, $K_2CO_3$, EtOH, reflux, 12hrs

**1”, 7”-Bis(4-cyanobiphenyl-4’-yl)heptane) CB7CB** (**7**): 3.48g (10.2 mmol) of (Heptane-1,7-diylbis(4,1-phenylene))diboronic acid (6), 4.10g (22.5 mmol) of 4-bromobenzonitrile and 7g (51.25 mmol) of potassium carbonate were suspended in a mixture of 100 mL ethanol and 20 mL water followed evacuation with vacuum. It was purged with argon and sonicated. Then 1.18g (1.02 mmol) of tetrakis(triphenylphoshine)palladium(0) was added to the above reaction mixture while argon is being purged continuously and heated to reflux for 12 hours. The solvent was evaporated and 60 mL dichloromethane and 60 mL water were subsequently added to the leftover residue. The organic phase separated was washed twice with water, brine, and dried over anhydrous $Na_2SO_4$. The solvent was evaporated and the crude product was purified by column chromatography on silica (100-200mesh) with dichloromethane as eluent. The solid compound was further purified by recrystallization using HPLC hexane – dichloromethane (9:1). A colorless (bright white) solid; yield 0.9g (55%); IR (KBr Pellet): $\nu_{max}$ in $cm^{-1}$: 2925, 2851, 2225, 1604, 1492, 1396, 1184 and 1004; UV-Vis: $\lambda_{max}$= 285nm, Ɛ = 3.93 x $10^2$ L $mol^{-1}$ $cm^{-1}$;$^1$H NMR (400 MHz, $CDCl_3$): δ 7.72 – 7.65 (m, 8H),7.51 (d, J = 8 Hz, 4H), 7.29-7.26 (d, J = 8 Hz, 4H), 2.68 (t, J = 8 Hz, 4H) and 1.67 (m, 4H), 1.54 (m, 6H);$^{13}$C NMR (100 MHz, $CDCl_3$) δ 145.4, 143.6, 136.4, 132.8, 132.5, 132.5, 129.1, 127.9, 127.45, 127.5, 127.0, 119, 110.5, 35.5, 31.3, 29.3 and 29.2. The above mentioned spectroscopic data obtained was found matching with the reported ones.

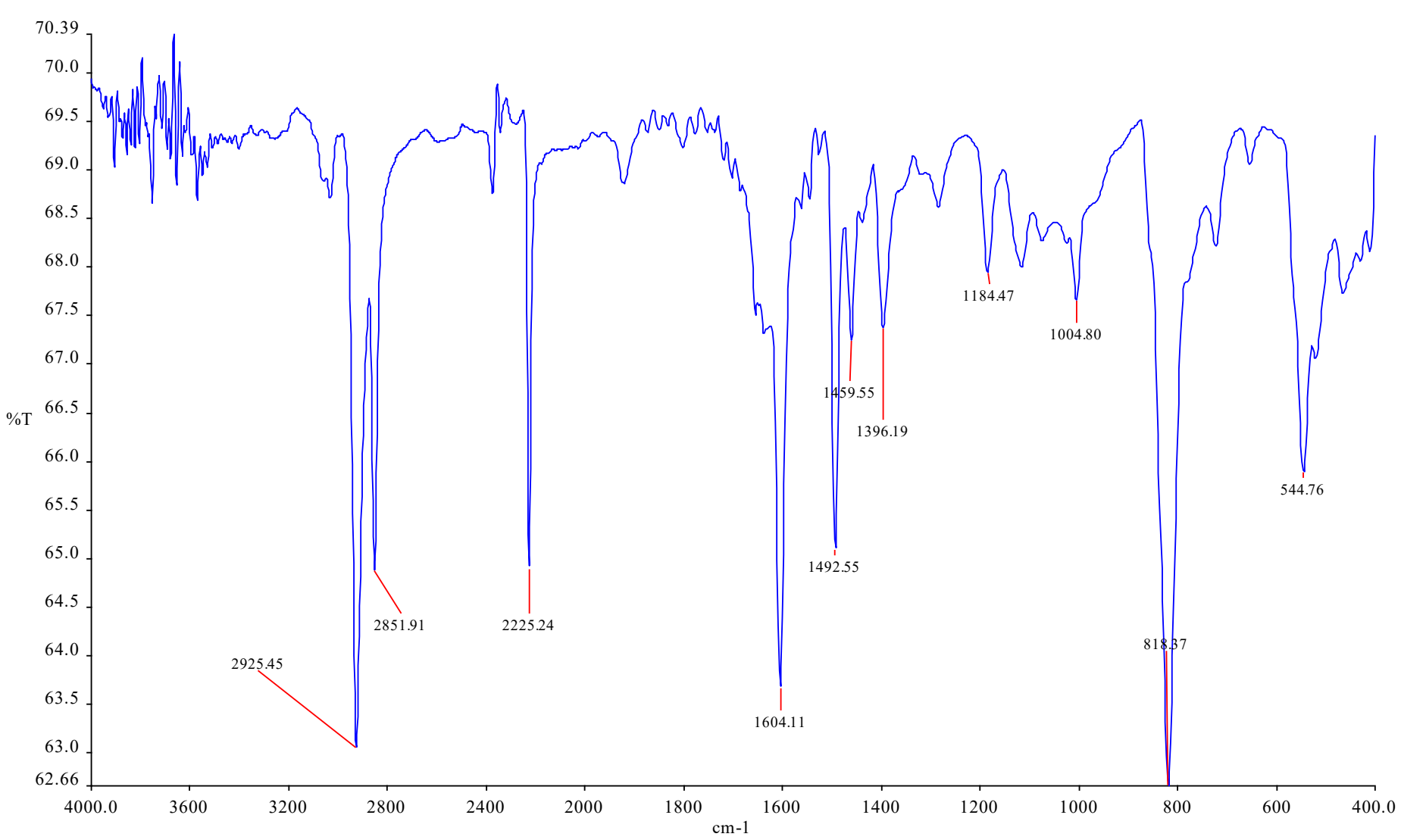

**Figure S1.** FTIR spectrum of **CB7CB**

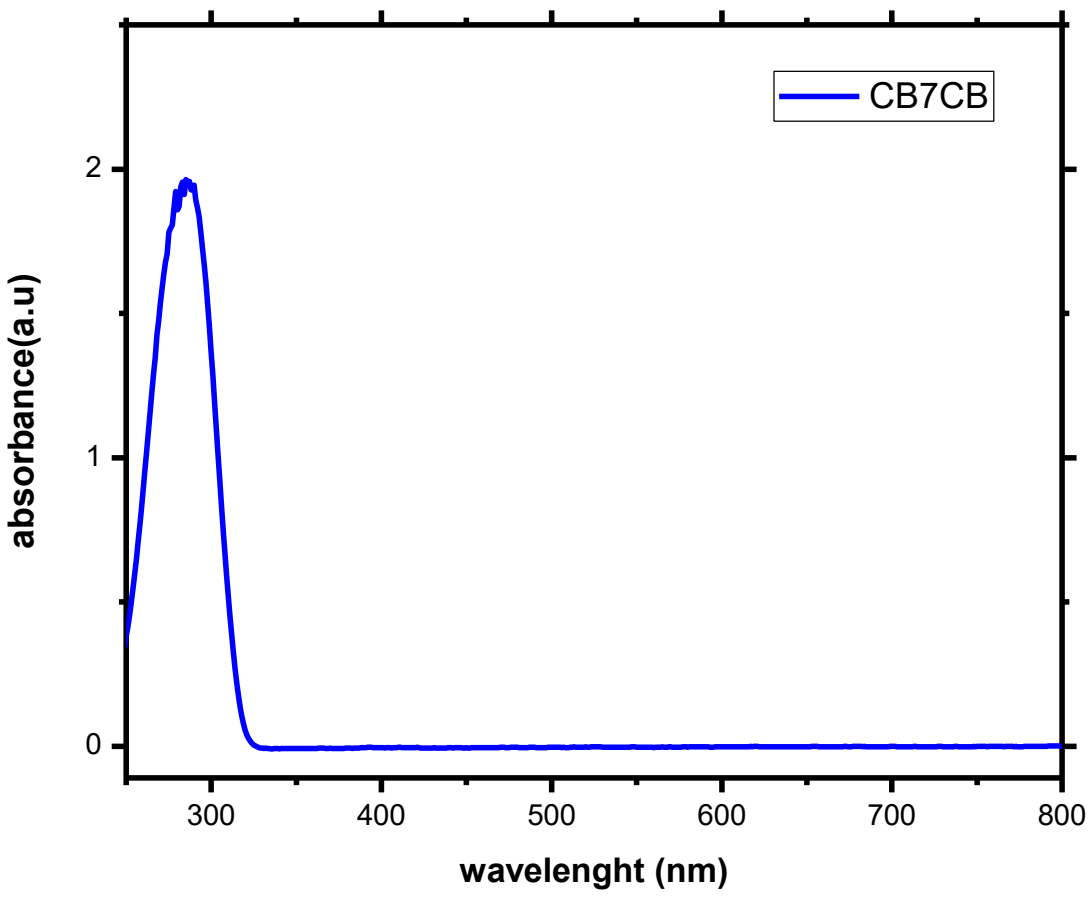

**Figure S2.** UV-Vis spectrum of **CB7CB** (DCM)

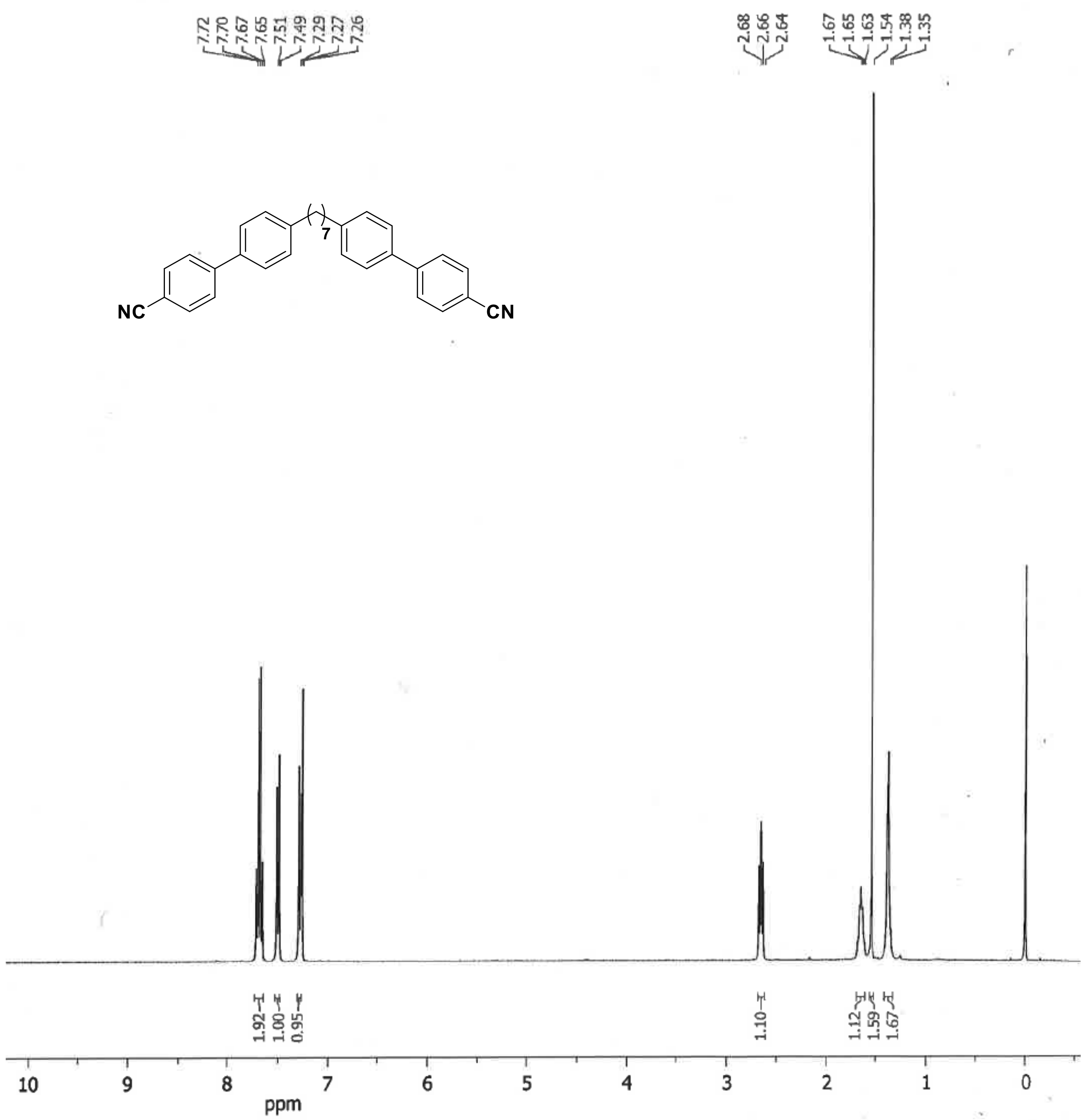

**Figure S3.**$^{1}$H-NMR spectra of **CB7CB** ($CDCl_3$, 400MHz)

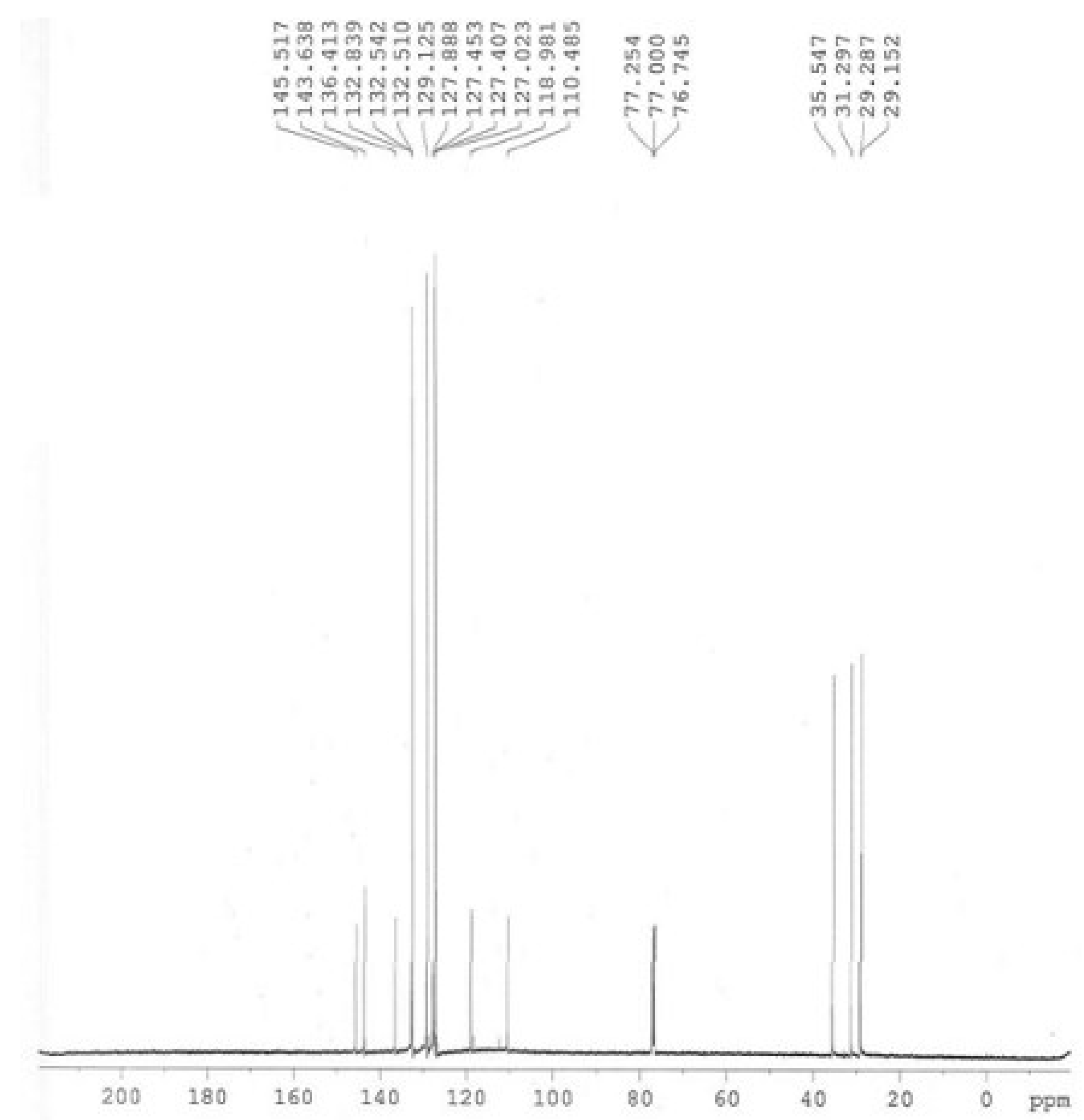

**Figure S4.** $^{13}$C NMR spectra of **CB7CB** ($CDCl_3$, 100 MHz)

**DSC Thermogram Traces:**

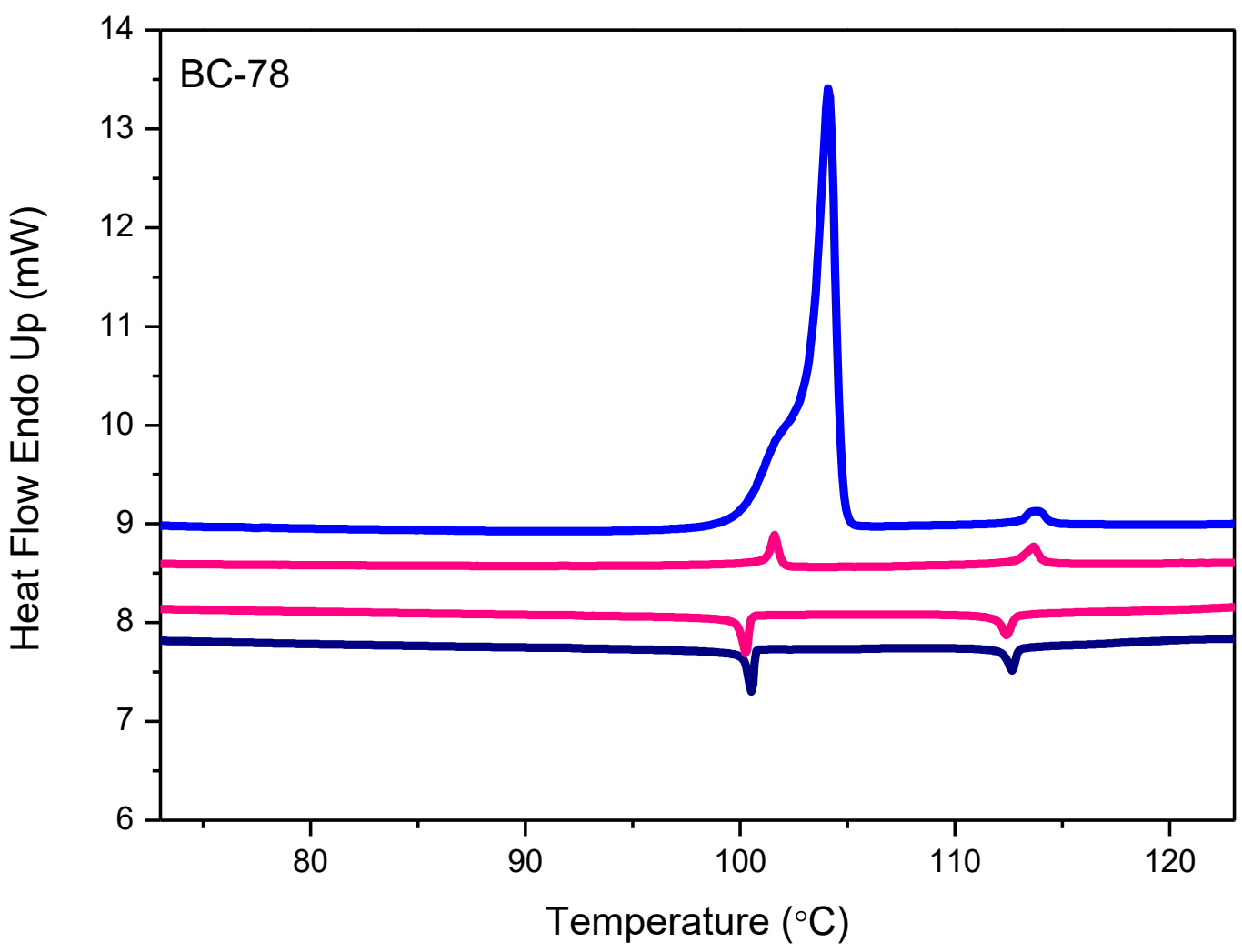

**Figure S5.** DSC traces recorded during the first and second heating-cooling cycles at a rate of 5 ºC/min. for CB7CB mesogens.

**Dielectric Spectroscopy:**

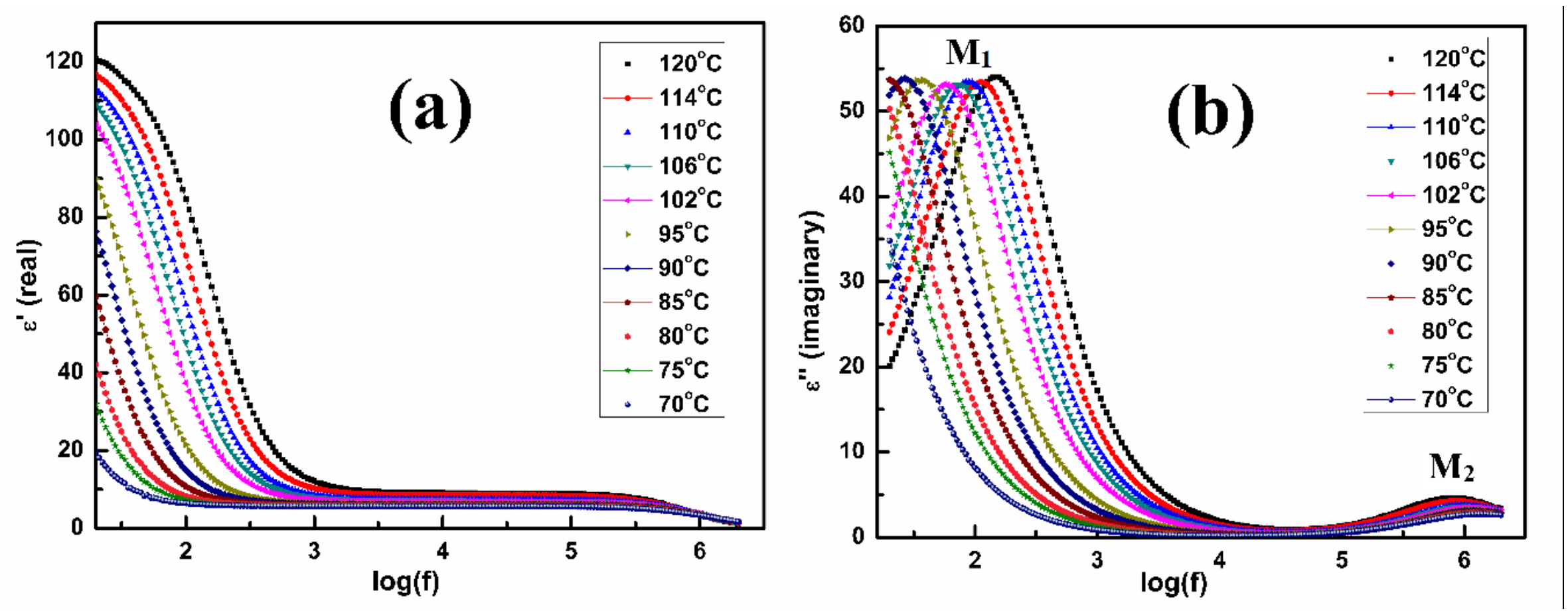

**Figure S6.** (a) Real (ε′) and (b) imaginary (ε'') part of dielectric permittivity of CB7CB as a function of frequency at different temperatures. Cell thickness: 5μm.

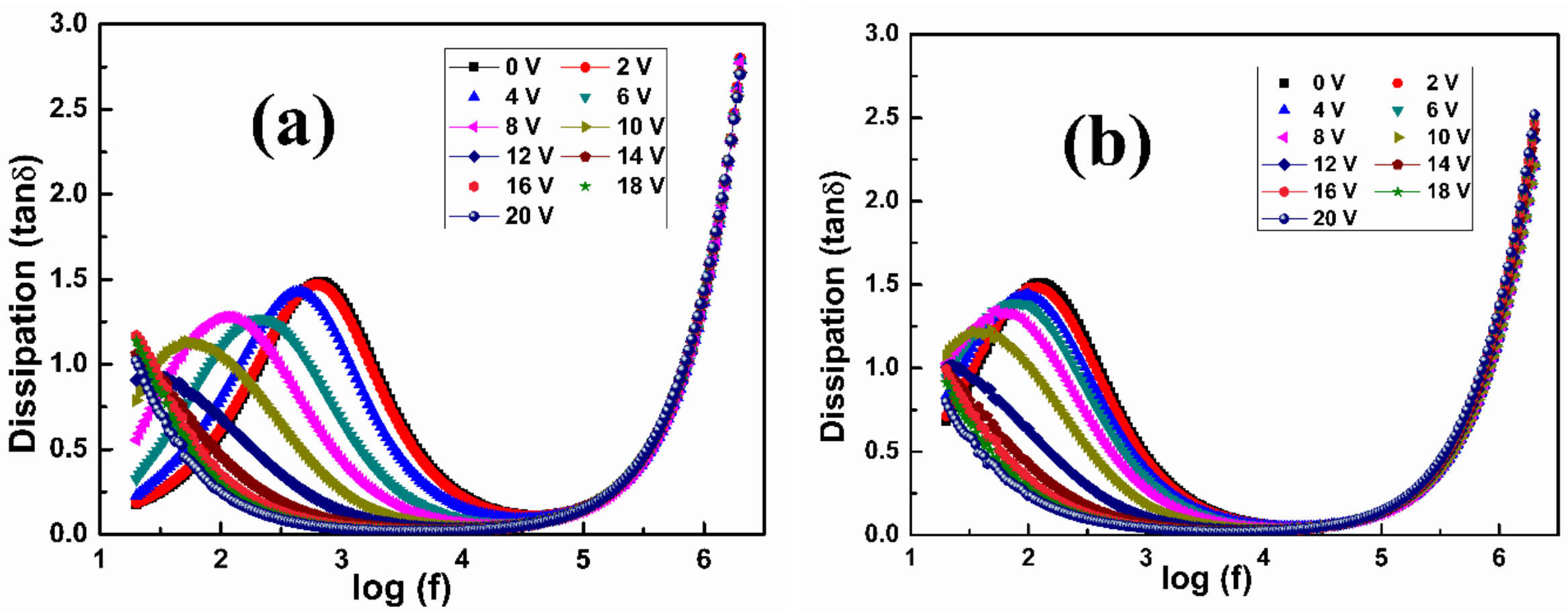

**Figure S7.** The dissipation factor (tan δ = ε''/ε') as a function of frequency at (a) 120 ºC and (b) 110 ºC on application of a DC bias voltage from 0 to 20 V. Cell thickness: 5μm.

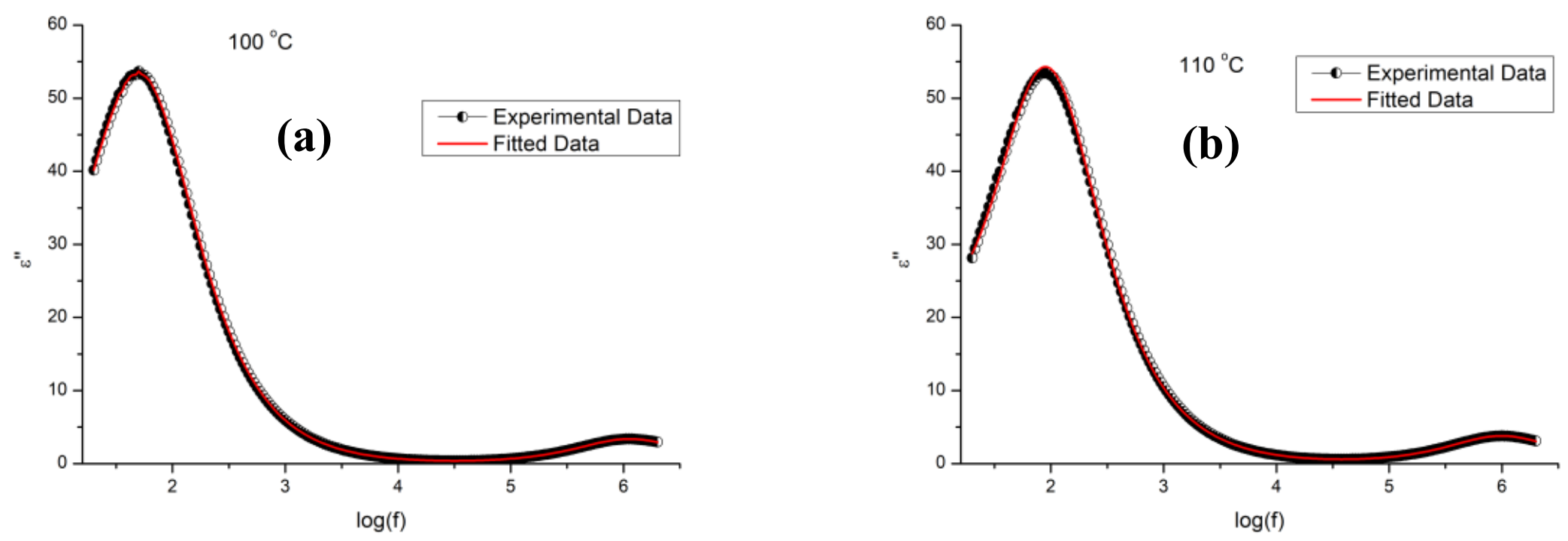

**Figure S8.** The experimental values of ε'' and the complete H-N fit to the experimental data at (a) 100 ºC and (b) 110 ºC as a function of frequency.

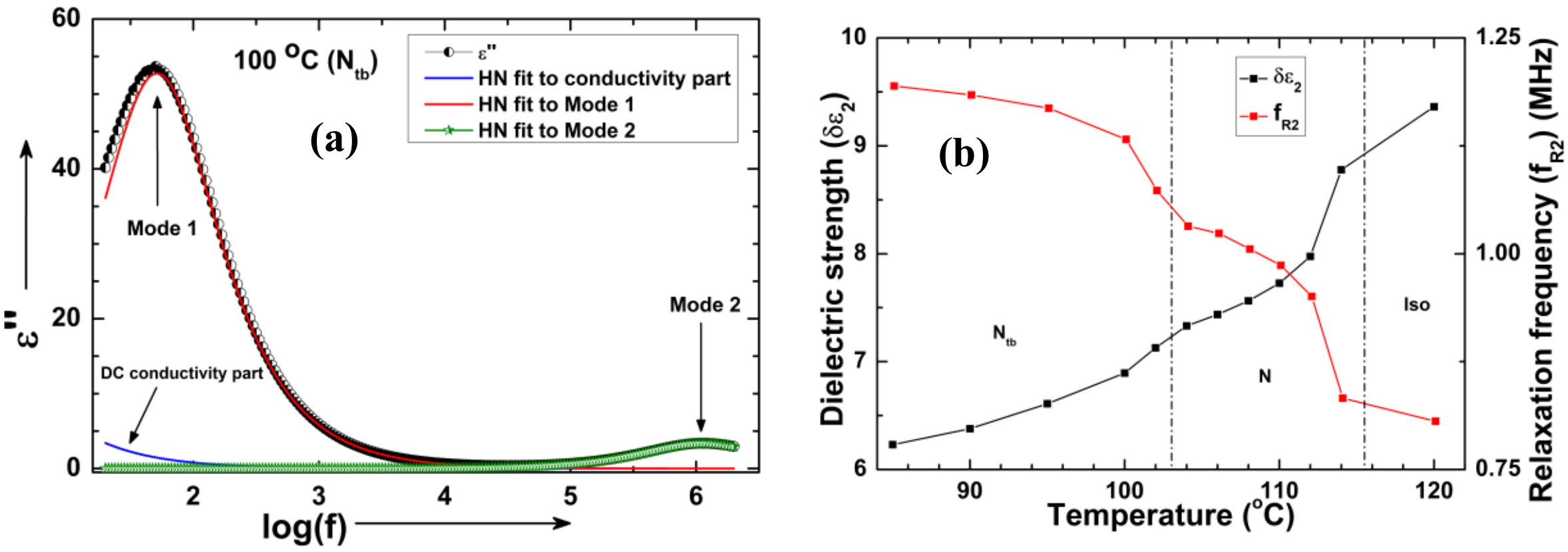

**Figure S9.** (a) The dielectric loss ($\varepsilon''$) as a function of frequency in the twist-bend nematic ($N_{tb}$) phase. The blue, red and green lines represent contributions from dc conductivity, mode $M_1$ and mode $M_2$), respectively, as obtained from H-N fitting; (b) Temperature dependence of the dielectric strength ($\delta\varepsilon$) and relaxation frequency ($f_R$) of mode $M_2$.

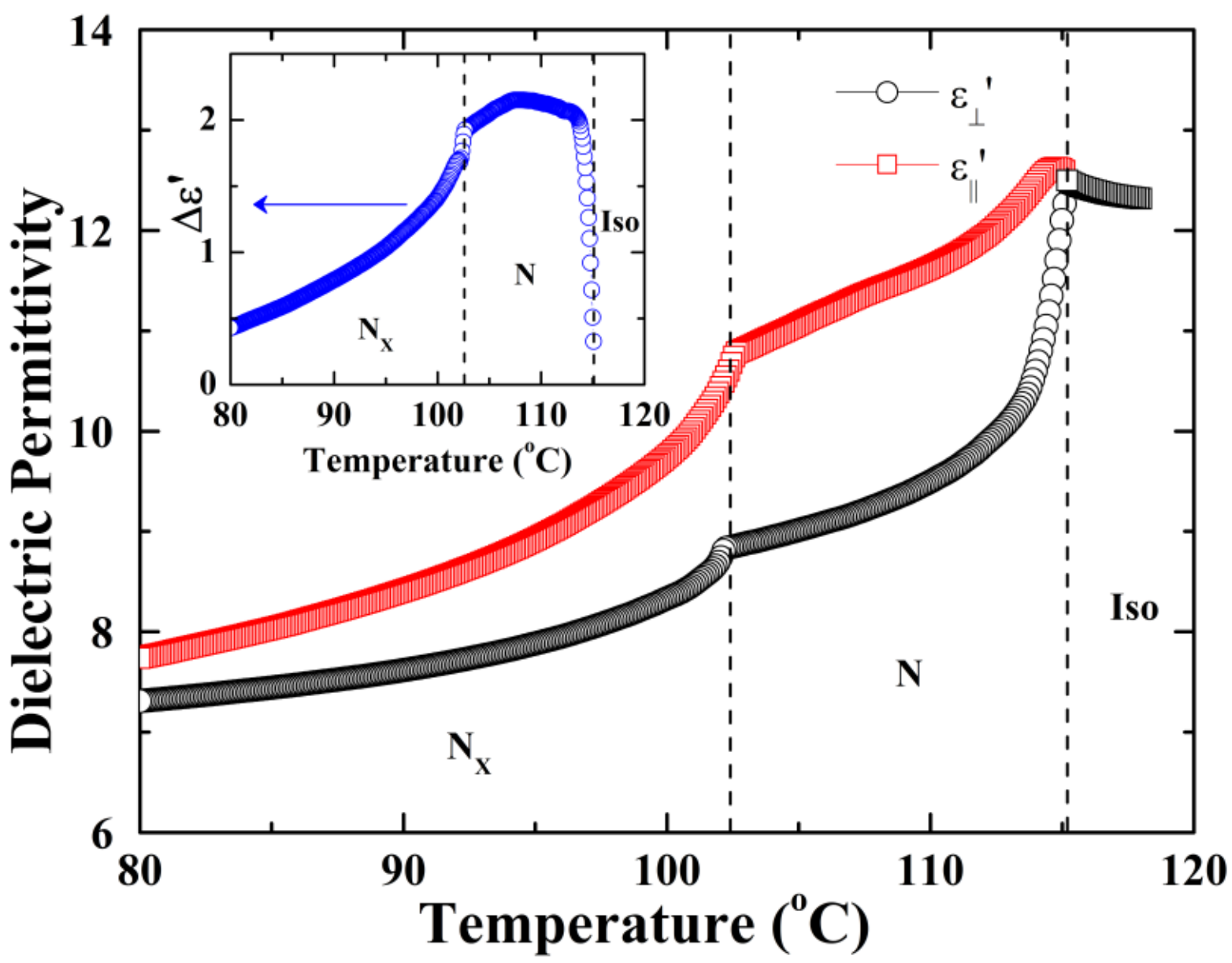

**Figure S10**. Static dielectric permittivity (1kHz), as a function of temperature. Inset: Dielectric anisotropy ($\Delta\varepsilon' = \varepsilon'_{||} - \varepsilon'_{\perp}$).

**Table T1. Comparative frequency-dependent dielectric studies reported in CB7CB and associated LC dimers showing $N_{tb}$ phase**

| Serial No. | Paper Details | | |
|---|---|---|---|
| 1. | Phase behavior and properties of the liquid-crystal dimer 1”,7”-bis(4-cyanobiphenyl-4’-yl) heptane: A twist-bend nematic liquid crystal – by **M. Cestari** and **G. R. Luckhurst** *et al.* Phys. Rev. E **84**, 031704 (2011) | | |
| | LC used: CB7CB | Cell: 50 μm - planar | Frequency Range: 100 Hz – 1.8 GHz (1 MHz - 1.8 GHz presented in paper) |
| | Modes observed: 2 modes (high frequency)<br>(i) mode **m1**: ~ 1 – 10 MHz – end-over-end or flip-flop motion of the dipolar groups.<br>(ii) mode **m2**: ~ 80-100 MHz – precessional motion of the dipolar groups. | | |
| 2. | Disentangling molecular motions involved in the glass transition of a twist-bend nematic liquid crystal through dielectric studies – by **D. O. López** and **G. R. Luckhurst** *et al.* J. Chem. Phys., **137**, 034502 (2012) | | |
| | LC used: CB7CB | Cell: 50 μm – (Gold-plated brass electrode cells) | Frequency Range: 0.01 Hz – 1.9 GHz |
| | Modes observed: 2 modes (high frequency)<br>(i) mode **m1**: ~ 1kHz – 10 MHz –flip-flop motion of the dipolar groups.<br>(ii) mode **m2**: ~ 10kHz - 100 MHz – precessional motion of the dipolar groups. | | |
| 3. | Soft modes of the dielectric response in the twist-bend nematic ($N_{TB}$) phase and identification of the transition to nematic splay bend $N_{SB}$ phase in the dimer CBC7CB – by **K. Merkel** and **G. H. Mehl** *et al.* arXiv: 1812.06838v1 / Phys. Chem. Chem. Phys., **21**, 22839 (2019). | | |
| | LC used: CB7CB | Cell: 1.6 μm - 10μm (varying) planar | Frequency Range: 0.1 Hz – 100 MHz |
| | Modes observed: 4 modes (3 high frequency and 1 low frequency) | | |

<table>
<tr><td></td><td colspan="3">(i) mode <b>m1</b>: ~ 8 MHz –flip-flop motion of the dipolar groups.<br>(ii) mode <b>m2</b>: ~ 70 MHz – precessional motion of the dipolar groups.<br>(iii) mode <b>m3</b>: 1 MHz – soft mode – associated with the fluctuations in tilt angle (θ). This mode m3 is quite similar to our high-frequency mode ($M_2$) in terms of $\delta\varepsilon$ and $f_R$ behaviour. This is associated with electroclinic and flexoelectric effects in the $N_{tb}$ phase similar to the N* phase.<br>(iv) mode <b>m4</b>: 1 Hz – 100 Hz – Identified as goldstone mode associated with the azimuthal angle (φ) fluctuations – This is similar to our low-frequency mode ($M_1$). Also an analogy for this can be drawn from the Debye type nature of our $M_1$ mode – pinned goldstone mode (D. Y. Kim <i>et al.</i>).<br>Further, they state that for E ~ 10V/μm, $N_{tb}$ – $N_{SB}$ transition may also take place – This is in line with the predictions by Pajak <i>et al.</i>, also a polar N phase can be realized with increasing field values.</td></tr>
<tr><td rowspan="3">4.</td><td colspan="3">Dielectric properties of liquid crystalline dimer mixtures exhibiting the nematic and twist-bend nematic phases – by <b>Nina Trbojevic</b> and <b>Mamatha Nagaraj</b> <i>et al.</i><br>Phys. Rev. E <b>96</b>, 052703 (2017)</td></tr>
<tr><td>LC used: CB7CB + 5CB(varying concentrations)</td><td>Cell: 10 μm planar</td><td>Frequency Range: 100 Hz – 2 MHz</td></tr>
<tr><td colspan="3">Modes observed: A single relaxation mode <b>m1</b>: 10 kHz – 100 kHz – associated with flip-flop motion of the dipolar groups. The dielectric strength ($\delta\varepsilon$) for m1 is also very small and the activation energy ($E_a$) is around 32 kJ/mol which is quite smaller than our $E_a$ for $M_1$ and quite larger than our $E_a$ for $M_2$.</td></tr>
<tr><td rowspan="3">5.</td><td colspan="3">Molecular dynamics of a binary mixture of twist-bend nematic liquid crystal dimers studied by dielectric spectroscopy – by <b>B. R. Herńandez</b> and <b>G. R. Luckhurst</b> <i>et al.</i><br>Phys. Rev. E <b>93</b>, 062705 (2016)</td></tr>
<tr><td>LC used: CB7CB + FFO9OCB (varying concentrations)</td><td>Cell: 8 μm planar</td><td>Frequency Range: 0.01 Hz – 1.8 GHz</td></tr>
<tr><td colspan="3">Modes observed: 2 modes (high frequency)<br>(i) mode <b>m1</b>: ~ 50kHz – 5 MHz – flip-flop/end-over-end motion of the dipolar groups.<br>(ii) mode <b>m2</b>: ~ 40MHz - 100 MHz – precessional motion of the dipolar groups.</td></tr>
<tr><td rowspan="3">6.</td><td colspan="3">Twist, tilt, and orientational order at the nematic to twist-bend nematic phase transition of CB9CB: A dielectric, 2H NMR, and calorimetric study – by <b>B. R. Herńandez</b> and <b>G. R. Luckhurst</b> <i>et al.</i><br>Phys. Rev. E <b>92</b>, 062505 (2015)</td></tr>
<tr><td>LC used: CB9CB</td><td>Cell: 50 μm</td><td>Frequency Range: 1 kHz – 1.8 GHz</td></tr>
<tr><td colspan="3">Modes observed: 2 modes (high frequency)<br>(i) mode <b>m1</b>: ~ 1-10 MHz – flip-flop/end-over-end motion of the dipolar groups.<br>(ii) mode <b>m2</b>: ~ 100 MHz – precessional motion of the dipolar groups.</td></tr>
<tr><td rowspan="3">7.</td><td colspan="3">Distortions in structures of the twist bend nematic phase of a bent-core liquid crystal by the electric field – by <b>K. Merkel</b>, <b>J. K. Vij</b> and <b>G. Shanker</b> <i>et al.</i><br>Phys. Rev. E <b>98</b>, 022704 (2018)</td></tr>
<tr><td>LC used: 5-Ring bent-core LC</td><td>Cell: 5 μm - planar</td><td>Frequency Range: 1 Hz – 100 MHz</td></tr>
<tr><td colspan="3">Modes observed: 4 modes (3 high frequency and 1 low frequency)<br>(i) mode <b>m1</b>: ~ 10 MHz – spinning motion of the dipolar groups.<br>(ii) mode <b>m2</b>: 30 kHz – 1 MHz - precessional motion of the dipolar groups.<br>(iii) mode <b>m3</b>: 10 - 20 kHz – tilt deformation mode. This is similar to our high frequency mode ($M_2$) where $f_R$ increases with decreasing temperature and corresponds to –ve activation energy.</td></tr>
</table>

<table>
<tr><td></td><td colspan="3">(iv) mode <b>m4</b>: 100 - 500 Hz – The collective structural mode associated with alternative compression and dilation of the pseudo-layered structure. This mode corresponds to a large polar order with δε ~ 800 and it is reminiscent of Goldstone mode observed in SmC* phase. This mode is very similar to our low-frequency mode $M_1$.</td></tr>
<tr><td rowspan="3">8.</td><td colspan="3">Dielectric response of electric-field distortions of the twist-bend nematic phase for LC dimers – by <b>K. Merkel, C. Welch, Z. Ahmed, W. Piecek and G. H. Mehl</b><br>J. Chem. Phys. <b>151</b>, 114908 (2019)</td></tr>
<tr><td>LC used: CB9CB and CB11CB</td><td>Cell: 5 µm - planar</td><td>Frequency Range: 1Hz – 100 Hz<br>1 kHz – 100 MHz</td></tr>
<tr><td colspan="3">Modes observed: 4 modes (3 high frequency and 1 low frequency)<br>The low frequency mode <b>m4</b>: 1 Hz- 100 Hz – Identified as the Goldstone mode related to long-scale fluctuations of the azimuthal angle – alternating compression and dilation of the pseudo-layered structure. The reasons for flexoelectric or electroconvective instabilities are ruled upo because they appear above certain threshold. The possibility of cybotactic clusters are also discussed but discarded since they did not observe any features, such as increasing rotational viscosity of increased cluster size on field application.</td></tr>
</table>

## Uemura Fitting:

To investigate the extent of ionic contributions to the obtained polarization responses, we have evaluated the concentration of free ions (*n*) and the diffusion coefficient (*D*). The values of *n* and *D* were calculated by fitting the modified Uemura equation [R3-R5] to the conductivity contributions (of $\varepsilon''$) extracted from H-N fitting.

$$\varepsilon'' = \left(\frac{nq^2D}{\omega\varepsilon_0 kT}\right)\left\{1+\frac{1-2e^A\sin(A)-e^{2A}}{A[1+2e^A\cos(A)+e^{2A}]}\right\} \qquad (1)$$

Here, *n* is the free ion concentration, *q* is the elementary charge, *D* is the diffusion coefficient, ω (= 2*πf*) is the angular frequency, *k* is the Boltzmann constant, *T* is the absolute temperature, $\varepsilon_0$ is the free-space permittivity and $A = d(\omega/2D)^{1/2}$ where *d* is the separation between electrodes.

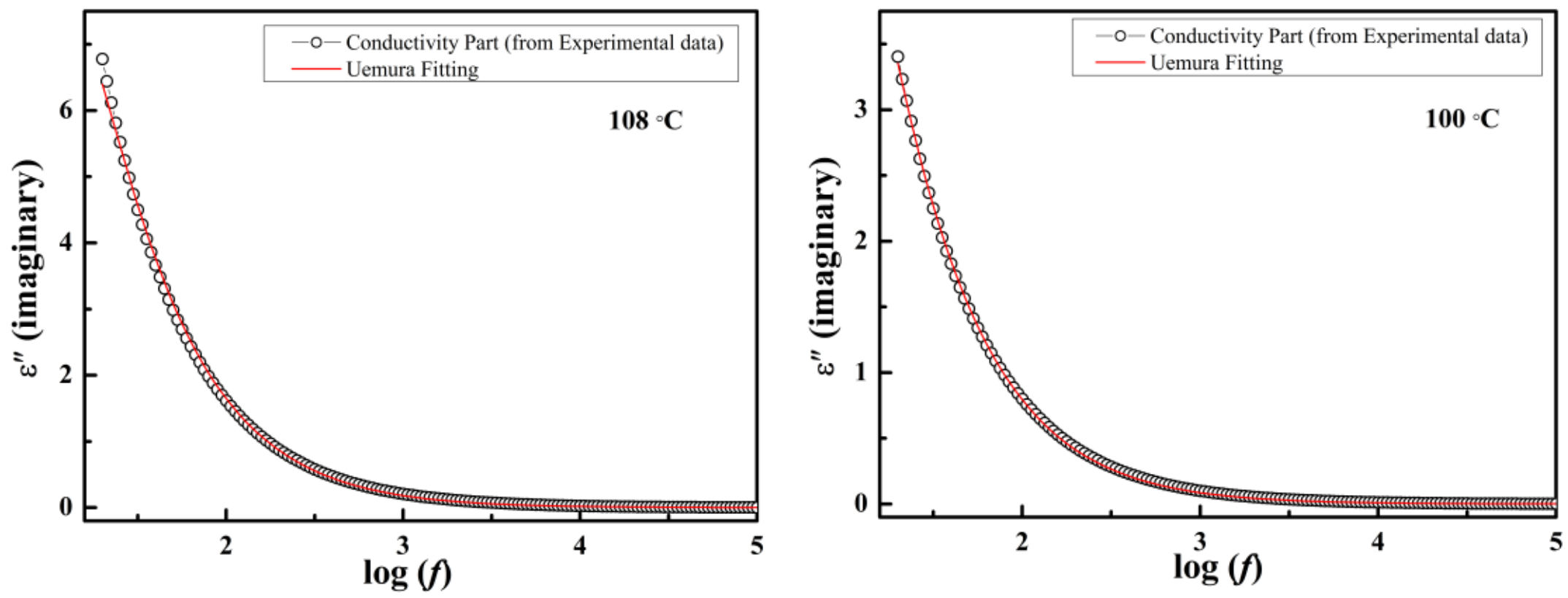

**Electro-optic responses:**

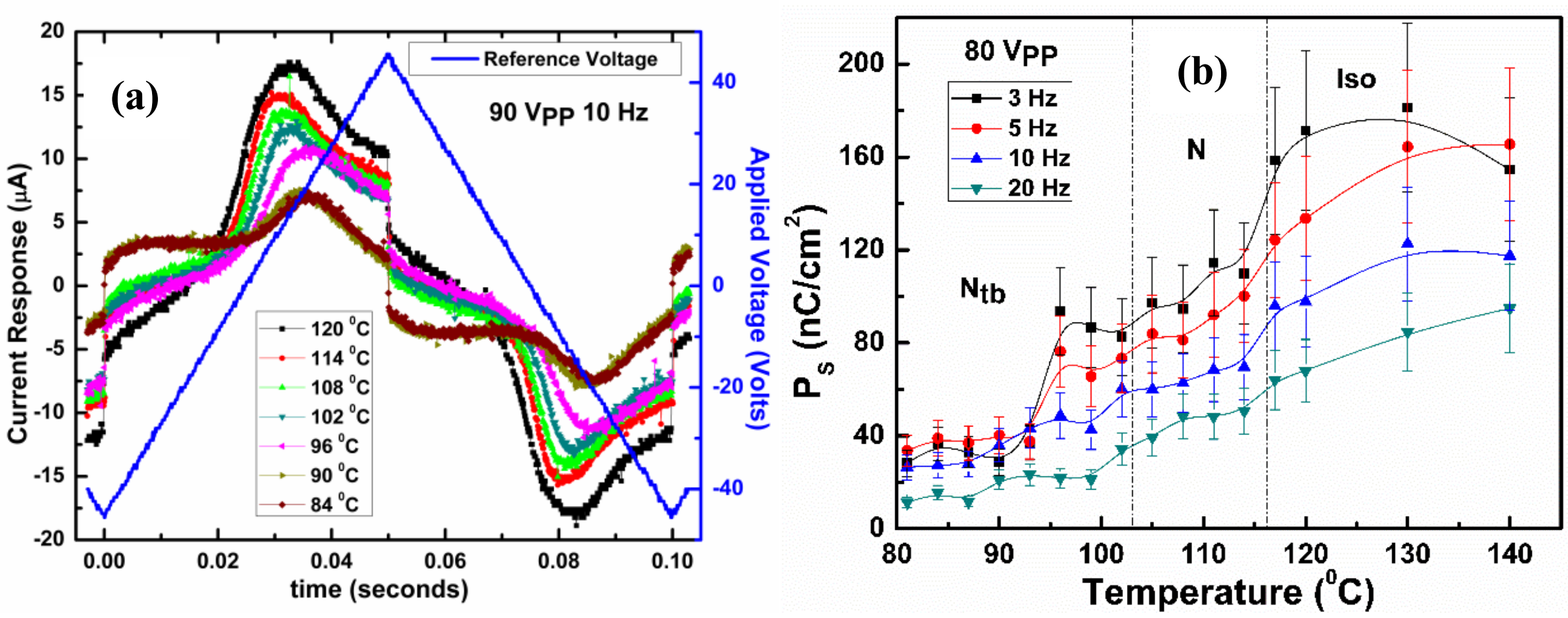

**Figure S11.** (a) Current response when $V_{Applied}$ = 90 $V_{PP}$ (18$V_{PP}$/μm) 10 Hz and (b) Spontaneous polarization ($P_S$) when $V_{Applied}$ = 80 $V_{PP}$ (16$V_{PP}$/μm) with varying frequency, at different temperatures. Cell thickness: 5μm, Input: Triangular voltage. Cell thickness: 5μm.

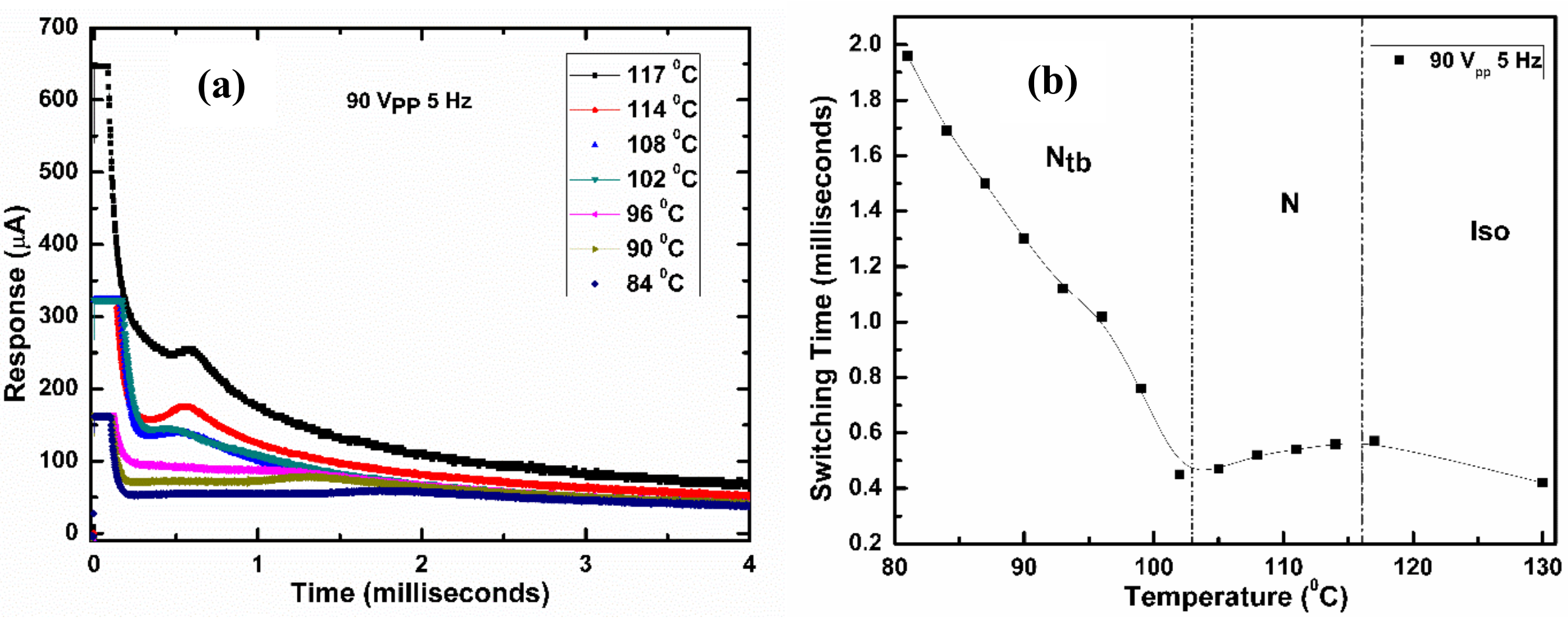

**Figure S12.** (a) The current response and (b) the associated switching time of the compound at different temperatures when subjected to a square-wave voltage of 80 $V_{PP}$ 5 Hz. Cell thickness: 5μm.

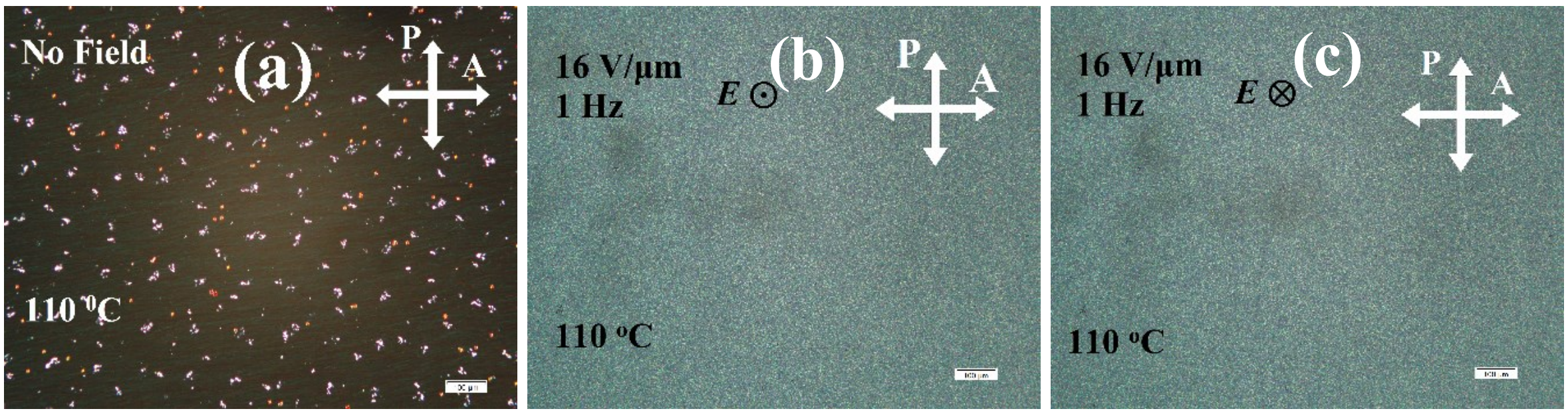

**Figure S13.** Textural switching observed in the sample at 110 °C (N): (a) No Field, (b-c) 80 $V_{PP}$ (16 V/μm) 1Hz. Cell thickness: 5μm.

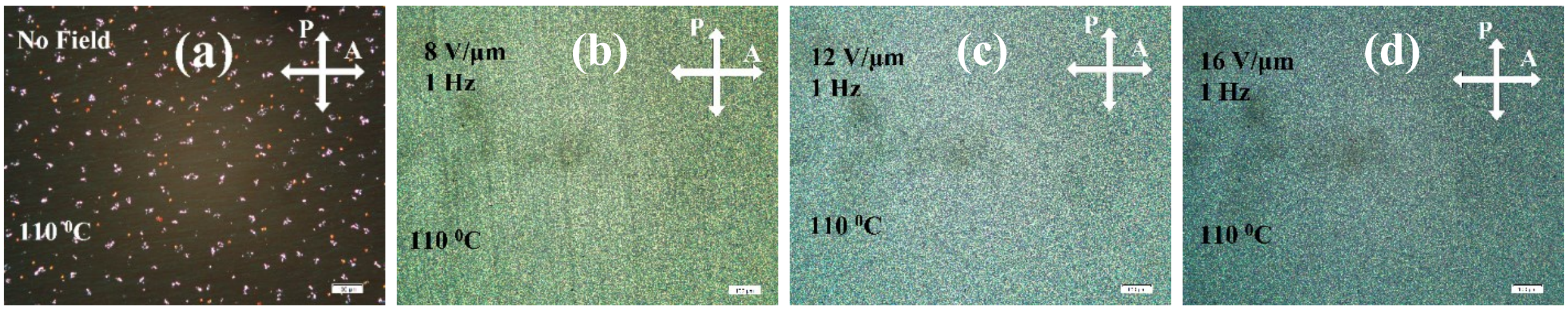

**Figure S14.** Textural switching observed in the sample at 110 °C (N): (a) No Field, (b) 40 $V_{PP}$ (8 V/μm) 1 Hz, (c) 60 $V_{PP}$ (12 V/μm) 1Hz and (d) 80 $V_{PP}$ (16 V/μm) 1 Hz. Cell thickness: 5μm.

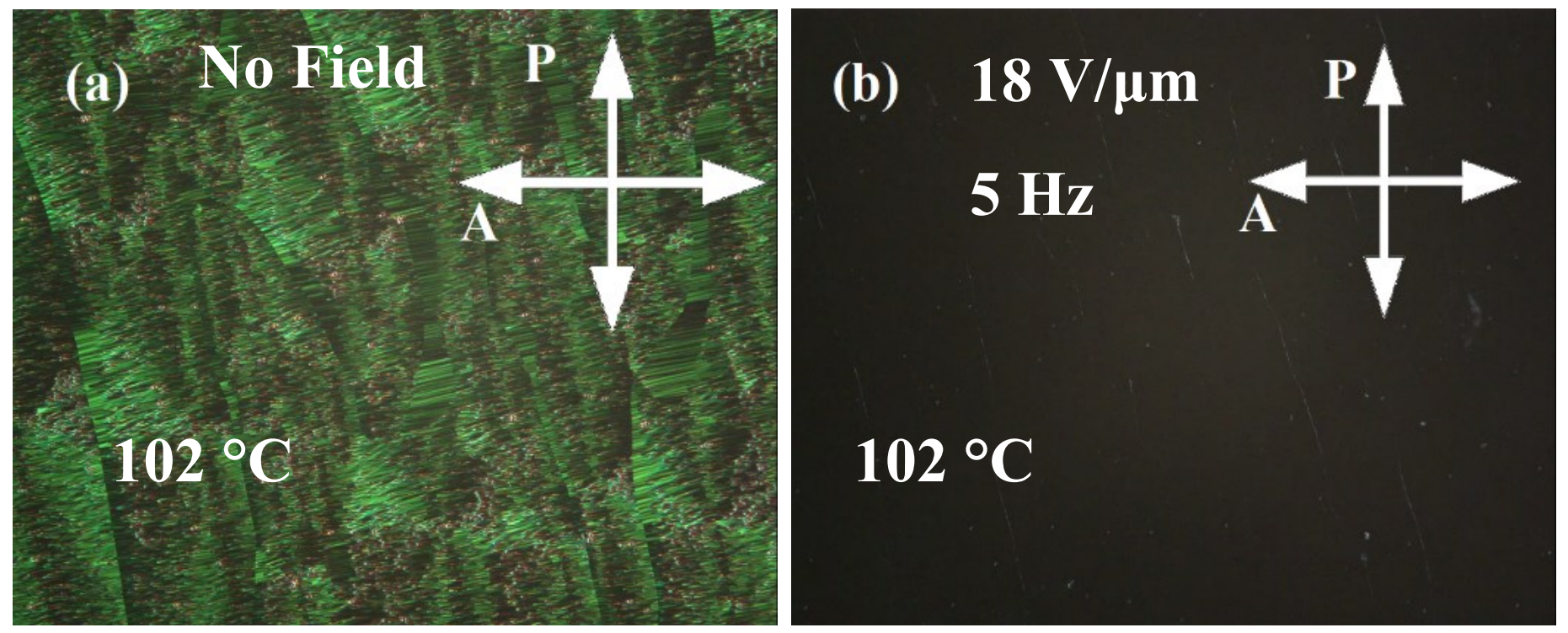

**Figure S15.** Unwinding of twist-bend helix under external electric-field at 102°C: (a) Rope-like pseudo focal-conic twist-bend texture when no field is applied (b) Dark texture when the twist-bend helix is unwound under external electric-field ~ 18 V/μm (f = 5 Hz). Cell thickness: 5μm.

**Optical transmission measurements:**

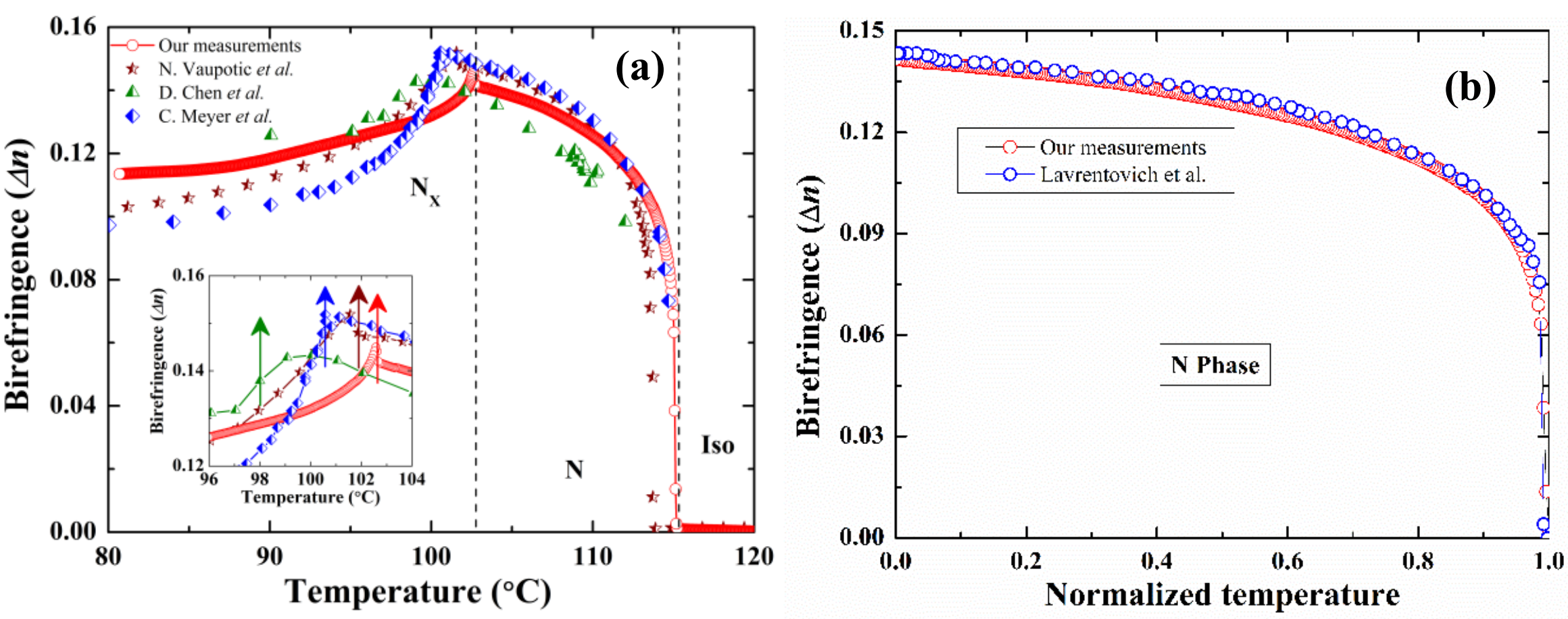

**Figure S16**. Comparison of our experimentally measured birefringence ($\Delta n$) data with the ones reported in literature – (a) with references [R2,R6-R7] and (b) in the N phase with reference [R8].

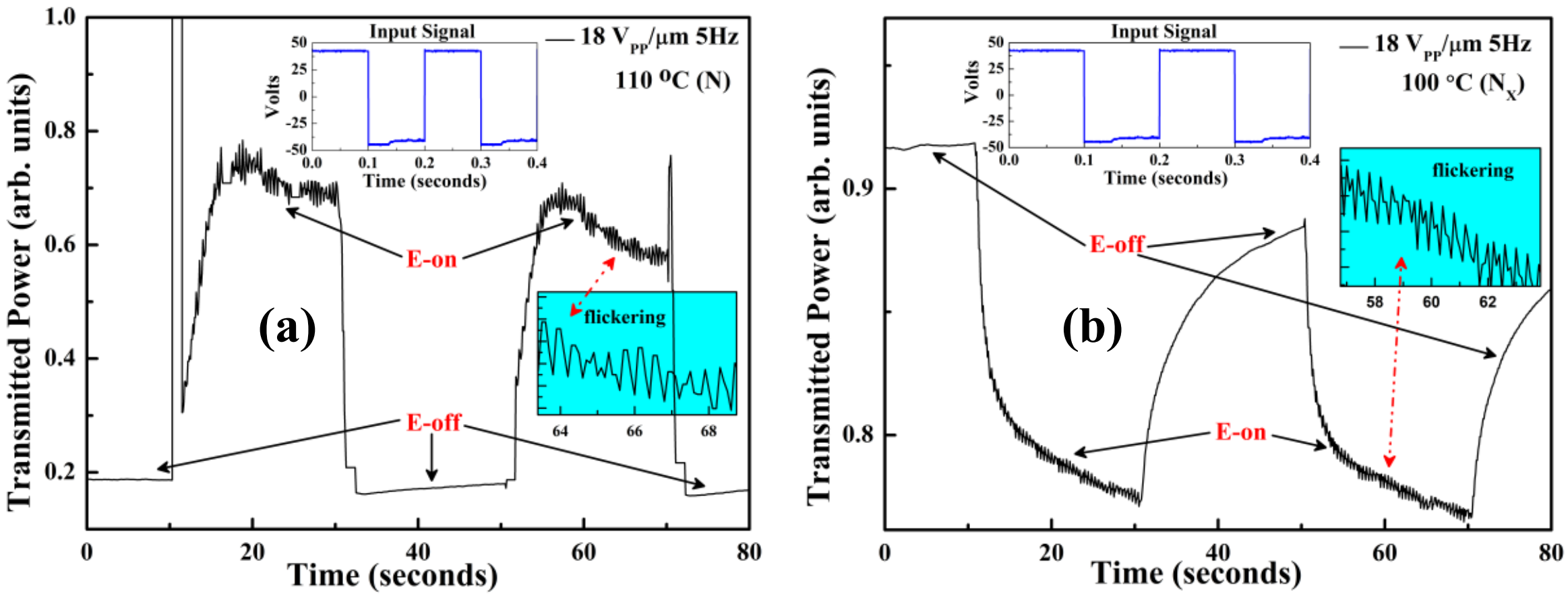

**Figure S17**. Time-dependent variation of transmission intensity for the LC CB7CB in (a) N phase and (b) $N_X$ phase. The input is an 18 $V_{PP}$/µm 5 Hz square-wave electric field (inset). The zoomed inset shows field-following flickering in the E-on state.

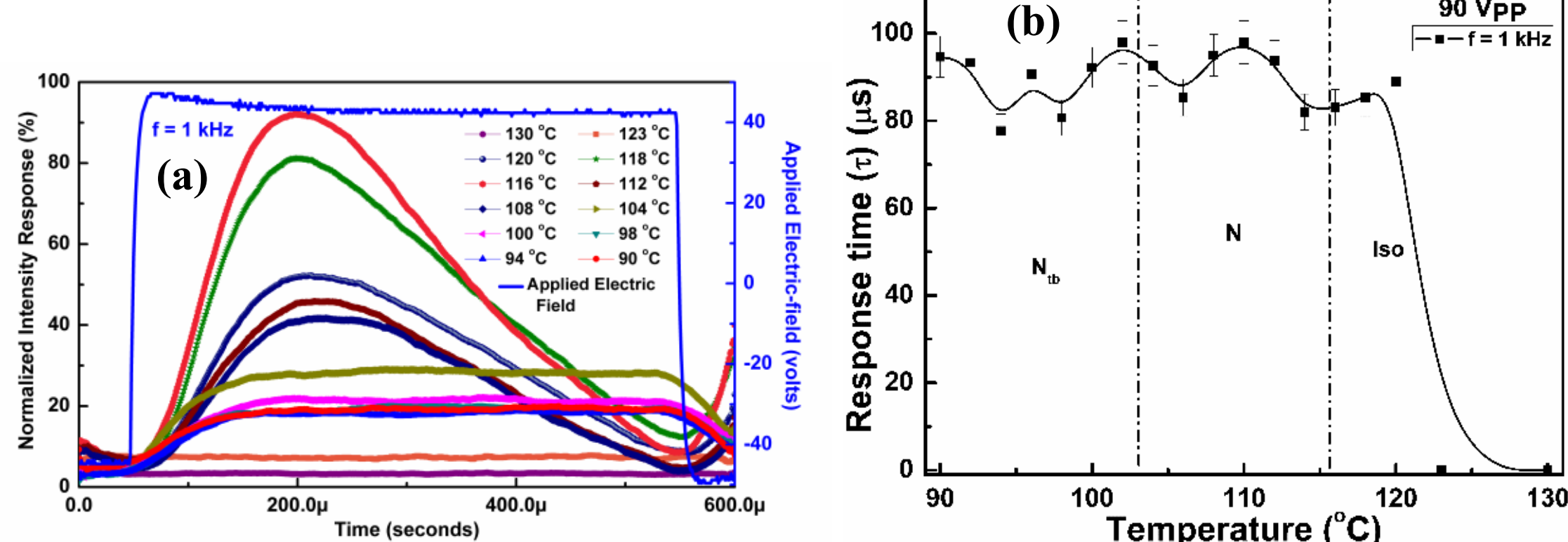

**Figure S18.** (a) Time dependant (smoothened) optical response signals (normalized) and (b) response time of the LC CB7CB at different temperatures. The input is a 90 $V_{PP}$ 1 kHz square-wave electric-field (indicated in blue in a). Cell thickness: 5µm.